\documentclass[12pt]{article}
\usepackage[dvipsnames]{xcolor}
\usepackage{mathrsfs,verbatim}
\usepackage{appendix}
\usepackage{natbib}
\usepackage{extarrows}
\usepackage{graphicx,setspace,lscape,longtable,epstopdf,xr}
\usepackage{mathrsfs,amsmath,amsthm,amssymb,color}
\usepackage{epsfig,graphicx}
\usepackage{rotating}
\usepackage{float}
\usepackage{bm}
\usepackage{ragged2e}
\usepackage{todonotes}
\usepackage{threeparttable}
\usepackage{multirow}
\usepackage{dsfont}
\usepackage{color}
\usepackage{xr}
\usepackage{booktabs}
\usepackage{algorithm}
\usepackage{algpseudocode}
\usepackage{amsmath}
\usepackage{graphicx}
\usepackage{indentfirst}
\usepackage{caption}
\usepackage{subcaption}
\usepackage[hidelinks, colorlinks=true,citecolor=blue]{hyperref}
\usepackage[T1]{fontenc}

\hypersetup{
colorlinks=true,
linkcolor=blue,
citecolor=blue,
urlcolor=blue}

\bibpunct{(}{)}{;}{a}{,}{,}

\newtheorem{theorem}{Theorem}
\newtheorem{lemma}{Lemma}
\newtheorem{proposition}{Proposition}
\newtheorem{definition}{Definition}
\newtheorem{assumption}{Assumption}
\def\one{{\bf 1}}
\def\bA{{\mathcal A}}
\def\ve{\varepsilon}

\def\beq{\begin{equation}}
\def\eeq{\end{equation}}
\def\beqr{\begin{eqnarray}}
\def\eeqr{\end{eqnarray}}
\def\beqrs{\begin{eqnarray*}}
\def\eeqrs{\end{eqnarray*}}
\def\bet{\begin{theorem}}
\def\eet{\end{theorem}}
\def\bel{\begin{lemma}}
\def\eel{\end{lemma}}
\def\bep{\begin{proposition}}
\def\eep{\end{proposition}}
\def\bg{\begin{figure}[tbph]\begin{center}}
\def\eg{\end{center}\end{figure}}

\def\bc{\begin{center}}
\def\ec{\end{center}}

\def\gap{\textup{gap}}

\newtheorem{remark}{Remark}
\newtheorem{corollary}{Corollary}

\def\wh{\widehat}

\def\E{\mathbb{E}}
\def\bB{\mathbf{B}}
\def\mA{\mathcal A}
\def\mC{\mathcal C}
\def\mN{\mathcal{N}}

\def\mG{\mathcal G}

\def\mH{\mathcal H}

\def\cR{\mathcal{R}}

\def\bY{\mathcal Y}

\def\bc{\mathbf c}

\def\mX{\mathbb{X}}
\def\cX{\mathcal{X}}

\def\mY{\mathbb{Y}}
\def\mZ{\mathbb{Z}}
\def\bx{\mathbf{x}}
\def\bZ{\mathbf{Z}}

\def\bv{\mathbf{v}}
\def\var{\mbox{var}}

\def\argmin{\mbox{argmin}}
\def\argmax{\mbox{argmax}}

\def\diag{\mbox{diag}}

\def\vec{\mbox{vec}}

\def\bgamma{\boldsymbol{\gamma}}
\def\bbeta{\boldsymbol{\beta}}

\def\bLambda{\boldsymbol{\Lambda}}
\def\bSigma{\boldsymbol{\Sigma}}

\newcommand{\RNum}[1]{\uppercase\expandafter{\romannumeral #1\relax}}
\def\bX{\mathbf{X}}
\def\bZ{\mathbf{Z}}

\def\by{\mathbf{y}}

\def\bY{\mathbf{Y}}
\def\bA{\mathbf{A}}

\def\bW{\mathbf{W}}

\def\b{\mathbf{b}}

\def\bg{\mbox{\boldmath $g$}}
\def\bI{\mathbf{I}}

\def\zero{\mathbf{0}}
\def\defeq{\stackrel{\mathrm{def}}{=}}  

\def\qic{\textup{QIC}}

\numberwithin{equation}{section}

\def\be{\begin{eqnarray*}}
\def\ese{\end{eqnarray*}}
\def\be{\begin{eqnarray}}
\def\ee{\end{eqnarray}}
\def\bsq{\begin{equation*}}
\def\esq{\end{equation*}}
\def\bq{\begin{equation}}
\def\eq{\end{equation}}

\def\var{\hbox{var}}

\def\wh{\widehat}

\def\mR{\mathbb{R}}

\def\vec{\mbox{vec}}
\def\argmin{\mbox{argmin}}
\def\argmax{\mbox{argmax}}
\def\diag{\mbox{diag}}

\def\trans{^{\top}}

\def\bTheta{\boldsymbol\Theta}

\def\beps{\boldsymbol\epsilon}

\def\bb{\boldsymbol\beta}

\def\bdelta{\boldsymbol\delta}

\def\bxi{\boldsymbol\xi}

\def\bzeta{\boldsymbol\zeta}

\def\bg{\boldsymbol\gamma}

\def\0{{\bf 0}}
\def\1{{\bf 1}}
\def\A{{\bf A}}

\def\G{{\bf G}}

\def\B{{\bf B}}

\def\D{{\bf D}}

\def\b{{\bf b}}
\def\I{{\bf I}}

\def\M{{\bf M}}
\def\bM{{\bf M}}

\def\L{{\bf L}}
\def\x{{\bf x}}
\def\I{{\bf I}}

\def\Y{{\bf Y}}

\def\C{{\bf C}}

\def\z{{\bf z}}

\def\diag{\hbox{diag}}

\def\bq{\begin{equation}}
\def\eq{\end{equation}}

\def\wh{\widehat}

\def\diag{\hbox{diag}}
\def\log{\hbox{log}}

\def\squarebox#1{\hbox to #1{\hfill\vbox to #1{\vfill}}}
\def\btheta{{\boldsymbol \theta}}
\def\bfeta{{\boldsymbol \eta}}
\def\balpha{{\boldsymbol \alpha}}

\def\blambda{{\boldsymbol \lambda}}

\def\bzeta{{\boldsymbol \zeta}}
\def\bpi{{\boldsymbol \pi}}
\def\bx{{\bf x}}
\def\bz{{\bf z}}
\def\vec{\mathrm{vec}}
\def\mA{\mathcal{A}}

\def\bE{\mathbf{E}}

\def\mC{\mathcal{C}}

\def\cS{\mathcal{S}}
\def\mH{\mathcal{H}}

\def\var{\hbox{var}}

\def\bse{\begin{eqnarray*} }
\def\ese{\end{eqnarray*}}
\def\be{\begin{eqnarray}}
\def\ee{\end{eqnarray}}
\def\bsq{\begin{equation*}}
\def\esq{\end{equation*}}
\def\bq{\begin{equation}}
\def\eq{\end{equation}}

\def\wh{\widehat}
\def\wt{\widetilde}
\def\diag{\hbox{diag}}

\def\diag{\hbox{diag}}

\def\o{\textup{or}}

\def\boxit#1{\vbox{\hrule\hbox{\vrule\kern6pt\vbox{\kern6pt#1\kern6pt}\kern6pt\vrule}\hrule}}

\begin{document}
\begin{center}

{\bf\Large  Matrix-valued Network Autoregression Model with Latent Group Structure}

\bigskip
		Yimeng Ren$^1$, Xuening Zhu$^{1*}$, Ganggang Xu$^{2*}$ and Yanyuan Ma$^3$

		{\it $^1$Fudan University, Shanghai, China;\\
$^2$University of Miami, USA;\\
            $^3$The Pennsylvania State University, PA, USA
}

\end{center}

\begin{singlespace}
		\begin{abstract}
Matrix-valued time series data are frequently observed in a
  broad range of areas and have attracted great attention
recently.
In this work, we model network effects for high
dimensional matrix-valued time series data in a matrix
autoregression framework.
To characterize the potential heterogeneity of the subjects
and handle the high dimensionality simultaneously, we assume
that each subject has a latent group label, which enables us to cluster
the subject into the corresponding row and column groups.
We propose a group matrix network autoregression (GMNAR) model, which
assumes that the subjects in the same group share the same set of
model parameters.
To estimate the model, we develop an iterative algorithm.
Theoretically, we show that the group-wise parameters and group
memberships can be consistently estimated when the group numbers are
correctly or possibly over-specified.
An information criterion for group number estimation is also provided
to consistently select the group numbers.
Lastly, we implement the method on a Yelp dataset to illustrate the
usefulness of the method.

			\vskip 1em
			\noindent {\bf KEY WORDS:}  Latent group;
                        Matrix-valued time series; Network data;
                        Vector autoregression.

		\end{abstract}
	\end{singlespace}

\begin{footnotetext}[1]{Xuening Zhu ({\it xueningzhu@fudan.edu.cn}) and Ganggang Xu are
    corresponding authors.}
\end{footnotetext}

\newpage

\section{Introduction}\label{sec:intro}

With the world becoming increasingly connected, studying network effects has become an important research topic in various disciplines, including economics, finance and many others. The primary focus of our study is to analyze the network effects inherent in high dimensional time series observed over multiple networks.  The existing literature has seen notable progress in the study of time series within a single network. For instance, \cite{zhu2017network} introduced a network autoregression model for investigating time series within large social networks,
while \cite{chen2023community} proposed a community network vector autoregression model for the high-dimensional time series.
In addition, \cite{ma2023sparse} delved into a sparse spatio-temporal autoregression model and estimated the model by profiling and bagging.
Recent contributions include \cite{jiang2023autoregressive}, which presented a first-order autoregressive model for dynamic network processes. In comparison to conventional high-dimensional time series models \citep{walden2002wavelet, leng2012sparse, zhou2014gemini, wang2019factor, chang2021modelling}, the network autoregressive model distinguishes itself by offering enhanced parameter interpretability, thus offering more insights into the intricacies of network dynamics.

In real-world scenarios, entities within a population commonly establish connections across multiple networks, often referred to as multi-relational or multi-layer networks. Recently, there has been significant interest in investigating these networks collectively, as seen in works like \cite{zhang2020flexible, jing2021community, ma2023community}.
While the majority of these studies focus on identifying community structures within multi-layer networks, there is also a considerable interest in quantifying the impacts of multi-relational network effects on various research objectives.  For example, \cite{emch2016integration} investigate the joint effects of spatial and social networks on disease transmission. \cite{chen2017uncovering} propose the utilization of network metrics from various social networks to predict the adoption of new products in marketing research. Lastly, \cite{corradini2021investigating} investigate the influence of multi-dimensional social networks on negative reviews posted on Yelp, among other similar studies.
Although these models are valuable in empirical research, a critical
gap remains in
the availability of rigorous statistical models with the capability of providing valid statistical inferences on multiple network effects, and it is our intention to address this gap.

\begin{figure}[htpb!]
	\begin{center}
		\includegraphics[width=0.7\textwidth]{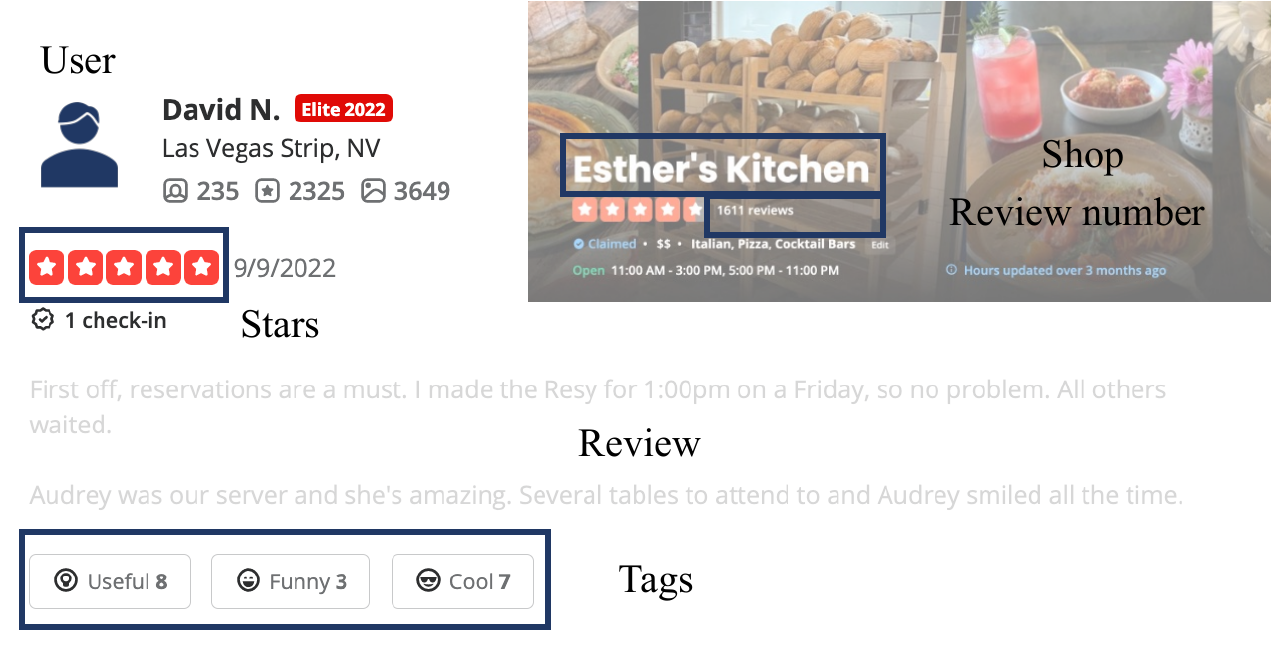}
		\caption{\small A review snapshot on the shop
			``Esther's Kitchen''. It contains the user
			information, shop statistics, review text and the
			tags assigned to this review.}
		\label{fig:yelp}
	\end{center}
\end{figure}

The central objective of our work is to investigate time series data indexed across multiple networks. To offer a more lucid representation of our methodology, we demonstrate it through a motivating dataset collected from Yelp (https://www.yelp.com/), which serves as a prominent review platform for various businesses, including restaurants, local retailers, entertainment establishments, and more. It also functions as a social platform where users can share information and their personal experiences.
The dataset covers the period from 2010 to 2018 and is collected from five North American cities (i.e., Charlotte, Las Vegas, Phoenix, Scottsdale and Toronto), and comprises four main categories of information: user data (e.g., user registration timestamps on Yelp), user-friend relationships, business information (including spatial location), and user reviews of businesses.
For example, Figure \ref{fig:yelp} illustrates a user's review of a restaurant named ``Esther's Kitchen'' in Las Vegas. In this case, the user gave the restaurant a five-star rating, and their review received 18 tags from other users, including 8 ``useful'', 3 ``funny'', and 7 ``cool'' tags. Overall, the restaurant has accumulated 1611 reviews.

\begin{figure}[htpb!]
	\begin{subfigure}[b]{0.3\textwidth}
		\centering
		\includegraphics[width=\textwidth]{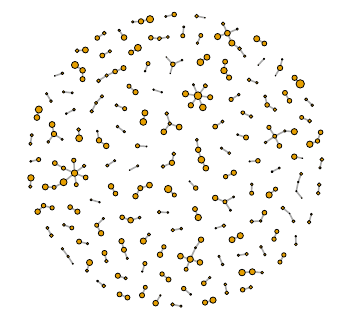}
		\caption{\small Social network}
		\label{fig:social_net}
	\end{subfigure}
	\begin{subfigure}[b]{0.3\textwidth}
		\centering
		\includegraphics[width=\textwidth]{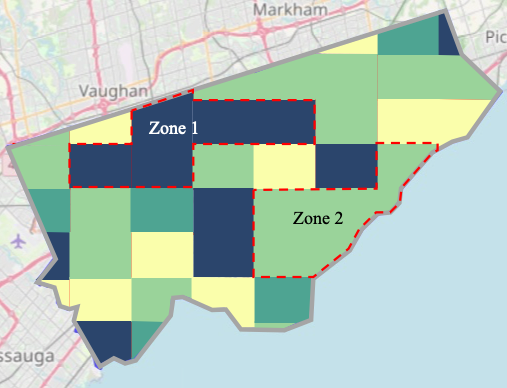}
		\caption{\small Spatial network}
		\label{fig:spat_net}
	\end{subfigure}
	\begin{subfigure}[b]{0.3\textwidth}
		\centering
		\includegraphics[width=\textwidth]{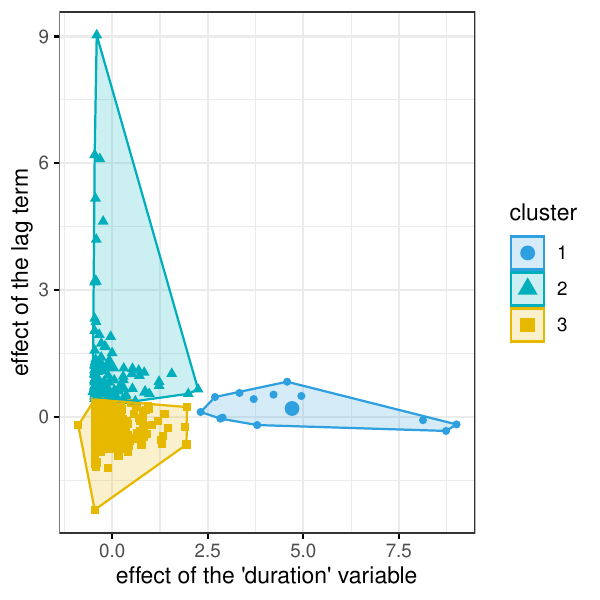}
		\caption{\small Estimated parameters}
		\label{fig:cluster_param}
	\end{subfigure}
	\caption{\small {(a) The social network of Toronto users with degrees greater than one. Node sizes reflect the logarithm of the number of comments made in 2008. (b) The spatial network of Toronto districts during the last quarter of 2018. Colors represent four quantile intervals of the logarithm of the number of comments made in the 4th quarter of 2008. Darker colors indicate districts with higher comment activity. (c) Clustering results based on estimated regression coefficients, categorizing users into three distinct groups. Each group is visually represented by different colors and shapes.}}
	\label{fig:pre_analysis}
\end{figure}

Within the Toronto segment of the dataset, we have records for $N_1=462$ users who have provided reviews for restaurants in $N_2=56$ distinct locations, spanning a timeframe of $T=36$ quarters.
Our primary focus in this analysis centers on the variable denoted as $Y_{ijt}$, representing the {$\log(1+x)$-transformed} number of reviews contributed by user $i$ to restaurants in district $j$ during the $t$th quarter. This variable forms a time series indexed by both the user ID ($i$) and the district ID ($j$).
The first challenge we encounter when analyzing this dataset is that neither users nor districts can be considered isolated units. As a result, it becomes imperative to model the $Y_{ijt}$'s in a collective manner. Specifically, users form a social network, while districts establish a spatial network that fosters substantial interactions among their respective network members. These interactions, in turn, significantly influence the outcome variable $Y_{ijt}$ when considered jointly. For example, \cite{tiwari2016social} find that peer social networks play a highly effective role in influencing restaurant preferences within social circles of friends. The social network analysis conducted on Yelp data by \cite{fe2023social} suggests that social network friends are 64\% more likely to visit the same restaurant when compared to non-friends. \cite{sun2017spatial} discovered significant spatial effects on ratings across various categories of Yelp venues and \cite{gan2021spatial} investigates the spatial network effects on the tourism economy. However, these studies are primarily empirical in nature and lack strong theoretical foundations in statistics. Furthermore, they tend to concentrate solely on the impact of a single network, rather than considering multiple network effects jointly.  For our motivating example, Figure \ref{fig:social_net} depicts the social network of users residing in Toronto who have at least two friends, highlighting the observation that connected friends often display similar comment volumes. Meanwhile, Figure \ref{fig:spat_net} presents the spatial network of Toronto, indicating that neighboring districts, such as zone 1 and 2, tend to exhibit similar  comment volumes. Both network effects contribute jointly to the outcome variable. Therefore, the first challenge we intend to address is how to construct a multivariate time series model that can rigorously quantify the impact of multiple network effects for data similar to our Yelp review dataset.

The second challenge we encounter in the Yelp review dataset relates to the heterogeneity among network members in both social and spatial networks. For instance, users with diverse posting habits and socioeconomic backgrounds may exhibit varying review patterns and distinct interactive relationships with their friends. Similarly, restaurants located in different spatial regions, such as the central business district or other areas, may experience different levels of popularity, leading to varying degrees of spatial spillover effects \citep[see, e.g.,][]{koschinsky2009spatial}.
For our motivating example, we conducted a preliminary regression analysis with the aggregated users' comment volumes (log transformed) as the response variable, denoted as $Y_{it} = \sum_{j} Y_{ijt}$. Two covariates were included: the lagged response ($Y_{i(t-1)}$) and the user's duration since registration (i.e., the number of months after joining Yelp by the $(t-1)$th quarter). The estimated regression coefficients obtained from all users were clustered into three groups using the $k$-means algorithm, and these clusters are visualized in Figure \ref{fig:cluster_param}. An evident heterogeneous pattern is observed among the users' coefficients. To tackle this challenge, we adopt the approach proposed in \cite{zhu2022simultaneous}, which involves dividing the members of both social and spatial networks into several sub-groups. We make the assumption that members within each sub-group share similar characteristics and behaviors.

Let $\bA_1 = (a_{1ij})\in \mR^{N_1\times N_1}$ and $\bA_2 = (a_{2ij})\in \mR^{N_2\times N_2}$ represent the adjacency matrices characterizing the social and spatial networks, respectively.
We assume the existence of $G$ groups in the social network and $H$ groups in the spatial network. The group membership for the $i$th node in the social network is denoted as $g_i$ ($1\le g_i\le G$), and the group membership for the $j$th node in the spatial network is denoted as $h_j$ ($1\le h_j\le H$). We propose the following model with a two-way group structure:
\be\label{eq:model0}
Y_{ijt} =\underbrace{\lambda_{g_i}\sum_{k = 1}^{N_1} \frac{a_{1ik}}{n_{1i}} Y_{kj(t-1)}}_{\text{\rm Social Network main effect}}
+
\underbrace{\gamma_{h_j}\sum_{k = 1}^{N_2} \frac{a_{2kj}}{n_{2j}}Y_{ik(t-1)}}_{\text{\rm Spatial Network main effect}}+ \underbrace{\alpha_{g_ih_j}Y_{ij(t-1)}}_{\text{\rm Self-momentum}}
+\underbrace{\bx_{it}\trans\bzeta_{g_i}+\bz_{jt}\trans\bdelta_{h_j}}_{\text{\rm Covariate effects}}+\ve_{ijt},
\ee
where $n_{1i}=\sum_{k=1}^{N_1}a_{1ik}$, $n_{2j}=\sum_{k=1}^{N_2}a_{2kj}$, $\bx_{it}\in \mR^{p_1}$ and $\bz_{jt}\in \mR^{p_2}$ are exogenous covariate vectors associated with the $i$th user and $j$th district, respectively, and $\ve_{ijt}$ represents independent and identically distributed (i.i.d) white noise with $E(\ve_{ijt})=0$ and $\var(\ve_{ijt}) = \sigma^2$. To ensure identifiability, it is required that $\sum_{g=1}^G\zeta_{g,1} = 0$ when both intercepts are included in $\x_{it}$ and $\z_{it}$, where $\zeta_{g_i,1}$ (the first element of $\bzeta_{g_i}$) represents the intercept for $\x_{it}$.

The first term in \eqref{eq:model0}, i.e., $\sum_{k = 1}^{N_1}
(a_{1ik}/n_{1i})Y_{kj(t-1)}$, represents the average  number of reviews ($\log(1+x)$-transformed)
by user $i$'s following friends on the restaurants in
district $j$ in the previous quarter. Consequently, $\lambda_{g_i}$
quantifies the influence of following friends on user $i$'s attitude
towards district $j$ and encapsulates a social network main effect. On
the other hand, the second term in \eqref{eq:model0}, i.e., $\sum_{k =
  1}^{N_2} (a_{2kj}/n_{2j}) Y_{ik(t-1)}$, calculates the average log
number of reviews by the $i$th user on districts that are connected with the $j$th district in the previous quarter. Thus, $\gamma_{h_j}$ signifies how user's  review count towards district $j$ is influenced by his/her reviews towards the districts connected with district $j$, and can be interpreted as the spatial network main effect.
Our empirical findings suggest that there exists different spatial effects within
different regions of the city.
Take Phoenix as an example, we find the regions in the central city
area possibly exhibit higher spatial effects, 
which indicates it is positively influenced by neighbouring regions for attracting customers.
Additionally, $\alpha_{g_ih_j}$ represents the self-driven time effect
for the $(i,j)$th time series, quantifying the momentum effect of the
review activity by the $i$th user towards the $j$th district in
the previous quarter. A higher value of $\alpha_{g_ih_j}$ suggests a
greater level of user loyalty or stickiness to the $j$th district. Finally, $\bzeta_{g_i}\in\mR^{p_1}$ and $\bdelta_{h_j}\in\mR^{p_2}$ are external covariate effects at the user and district levels, enhancing the model's capacity to account for user and district heterogeneity in the data. By evaluating the significance of $\lambda_{g}$'s and $\gamma_h$'s, one can rigorously examine the presence of social and spatial network main effects while controlling for other factors in Model~\eqref{eq:model0}.

Although Model~\eqref{eq:model0} can be readily extended to incorporate more than two networks (see Section~\ref{sec:conclusion} for details), for the sake of simplicity in presentation, we will focus on the case involving only two networks in this work. In this scenario, the observed response variable $\bY_t = (Y_{ijt})\in \mR^{N_1\times N_2}$ can be viewed as a matrix-valued time series with a two-way group structure. As a result, we refer to Model~\eqref{eq:model0} as the Group Matrix Network Autoregression (GMNAR) model. Our main contributions can be summarized as follows.
First, we introduce a highly interpretable network autoregression model for high-dimensional multivariate time series indexed by multiple networks. Second, to account for network heterogeneity, we impose a two-way group structure on related networks. Third, we establish estimation consistency for both model parameters and two-way group memberships, even when the numbers of groups are either correctly specified or over-specified. Lastly, we develop a group selection criterion to consistently determine the true group numbers and establish asymptotic normality when the group numbers are correctly specified. This theoretical framework enables rigorous tests for multiple network effects, which are crucial in various research disciplines.

The remainder of this article is structured as follows. In Section \ref{sec:model}, we introduce the notations used throughout the paper and present the Group Matrix Network Autoregression model. Section \ref{sec:estimate} outlines the model estimation procedure and the method for selecting the appropriate number of groups. Theoretical properties concerning parameter estimation, group membership estimation, and the estimation of the number of groups are discussed in Section \ref{sec:theory}. In Section \ref{sec:simulation}, we present a series of simulation experiments to illustrate the finite sample performance of our proposed method. Section \ref{sec:real} includes an application of our method to the Yelp dataset. Finally, Section \ref{sec:conclusion} provides concluding remarks.
Additional technical details and proofs, several numerical results, as well as another application on the trading data can be found in the supplementary material.

 \section{Matrix Network Autoregression with a Two-way Group Structure}

\subsection{General Model and Notations}\label{sec:model}
Consider a matrix-valued time series $\bY_t = (Y_{ijt})\in \mR^{N_1\times N_2}$ collected from two groups of subjects: $N_1$ row subjects and $N_2$ column subjects. We assume the presence of network structures for both the matrix rows and columns, characterized by a row adjacency matrix $\bA_1 = (a_{1ij})\in \mR^{N_1\times N_1}$ and a column adjacency matrix $\bA_2 = (a_{2ij})\in \mR^{N_2\times N_2}$. For instance, $\bA_1$ may represent the social network relationships among users, where $a_{1ij} = 1$ implies that the $i$th user follows the $j$th user, and $a_{1ij} = 0$ otherwise. Similarly, $\bA_2$ may reflect the spatial adjacency relationships among locations, where $a_{2ij} = 1$ indicates that the $i$th location is a spatial neighbor of the $j$th location, and $a_{2ij} = 0$ otherwise. We follow the convention of setting $a_{1ii} = 0$ for $1\le i\le N_1$ and $a_{2jj} = 0$ for $1\le j\le N_2$. Let's define $\bW_1 = (a_{1ij}/n_{1i})$ and $\bW_2 = (a_{2ij}/n_{2j})$ as the row and column-normalized adjacency matrices of $\bA_1$ and $\bA_2$, respectively. Here, $n_{1i} = \sum_{k=1}^{N_1} a_{1ik}$ and $n_{2j} = \sum_{k=1}^{N_2} a_{2kj}$.

We aim to model the dynamics of $\bY_t$ while incorporating group structures for both the row and column subjects. We assume the existence of $G$ row groups and $H$ column groups. For each row subject indexed by $i$, we denote its membership as $g_i$ ($1\le g_i\le G$), and for each column subject indexed by $j$, we denote its membership as $h_j$ ($1\le h_j\le H$). Using matrix notation, the GMNAR model~(\ref{eq:model0}) can be expressed as follows:
\be\label{eq:model}
\bY_t = \L \bW_1\bY_{t-1} + \bY_{t-1}\bW_2\G + \A\circ
\bY_{t-1} +\bb_{X, t}\1_{N_2}\trans+
\1_{N_1}\bb_{Z, t}^\top+\bE_t,
\ee
where
$\L = \diag(\lambda_{g_i}:1\le i\le N_1) \in \mR^{N_1 \times N_1}$,
$\G = \diag(\gamma_{h_j}:1\le j\le N_2) \in \mR^{N_2 \times N_2}$,
$\A = (\alpha_{g_ih_j}:1\le i\le N_1, 1\le j\le N_2)  \in \mR^{N_1 \times N_2}$,
$\bbeta_{X, t} = (\bx_{it}^\top \bzeta_{g_i}:1\le i\le N_1)^\top  \in \mR^{N_1}$,
$\bbeta_{Z, t} = (\bz_{jt}^\top \bdelta_{h_j}:1\le j\le N_2)^\top \in \mR^{N_2}$,
and $\bE_t = (\ve_{ijt})  \in \mR^{N_1 \times N_2}$. Here we use $\A\circ\B$ to denote the
hadamard product between matrices $\A$ and $\B$.

Throughout the paper, we use the following notations.
Denote $[n] = \{1,2,\cdots ,n\}$ for an integer $n$.
For a matrix $\bM = (m_{ij})\in \mR^{n_1\times n_2}$, let
$\bM_{i\cdot}$ be the $i$th row vector and $\bM_{\cdot j}$ as the
$j$th column vector of $\bM$.
In addition, let $\bM^{(\mC,\cdot)} = (m_{ij}: i\in \mC, j\in [n_2])$
and $\bM^{(\cdot,\mC)} = (m_{ij}: i\in [n_1], j\in \mC)$,
where $\mC$ is an index set.
Denote $\A\circ\B \in \mR^{n_1 \times n_2}$ as the
Hadamard product between matrices $\A\in \mR^{n_1\times n_2}$ and
$\B\in \mR^{n_1\times n_2}$.
For a symmetric matrix $\bM$, define
$\lambda_{\min}(\bM)$ and $\lambda_{\max}(\bM)$
as the corresponding smallest and largest eigenvalues.
For a vector, matrix, or tensor $\bM$, let $\|\bM\|_{\max}$
denote its largest absolute entry. For a vector $\bv = (v_j: j\in [p])^\top\in \mR^p$, let $\|\bv\| = (\sum_{j = 1}^p v_j^2)^{1/2}$.
For a set $\cS$, denote $|\cS|$ as the cardinal number of $\cS$.
Define $a_N \gg b_N$ as $a_N / b_N \rightarrow \infty$ as $N \rightarrow \infty$.
Moreover, denote $\one_p$ as a $p$-dimensional vector with all elements equal to one.
Denote $\I_p$ as an identity matrix with dimension $p\times p$.
For the group information, define $\mG = (g_i:1\le i\le N_1)^\top \in \mR^{N_1}$
	and $\mH = ({ h}_j: 1\le j\le N_2)^\top \in \mR^{N_2}$ as the row and column membership vectors, with the true vectors as $\mG^0$ and $\mH^0$.
	Let $\cR_g = \{i:g_i = g\}$ and $ \mC_h = \{j:h_j = h\}$, and further denote $N_{1g} = |\cR_g|$ and $N_{2h} = |\mC_h|$.
Accordingly, let $\cR_g^0 = \{i:g_i^0 = g\}$ and $ \mC_h^0 = \{j:h_j^0 = h\}$ be the memberships defined with true memberships $\{g_i^0\}$ and $\{h_j^0\}$.

{
\subsection{Comparisons with Existing Literature}\label{subsec:compare_literature}

{
The utilization of latent group structures to model heterogeneous data
has a well-established history in panel data analysis. For instance,
\cite{bonhomme2015grouped} and \cite{bester2016grouped} introduced
grouped linear panel models with time-varying fixed effects and
individual fixed effects, respectively. \cite{su2016identifying}
introduced a Classifier Lasso (C-Lasso) estimator for panel models,
and more recently, \cite{liu2020identification} explored estimation
and inference in cases of possible over-specification of the group
number. There have also been recent efforts to leverage latent group
structures in modeling time series data within a single network, as
demonstrated by
\cite{zhu2020grouped,zhu2022simultaneous, chen2023community}. However,
these works have predominantly focused on time series observed on
a single network, and to the best of our knowledge, our work is the first to tackle time series indexed by multiple networks with distinct group structures. While our primary presentation centers on the GMNAR model with two networks, the extension to multiple networks is straightforward (see Section~\ref{sec:conclusion}). This extension provides a novel auto-regression framework that dynamically constructs input features based on various network structures, facilitating statistical inferences on different types of network effects. Moreover, addressing the theoretical challenges involved in extending from a single network to multiple networks is non-trivial due to the interactions between different group structures, necessitating the development of new theoretical tools. In summary, our proposed framework equips researchers with valuable tools for analyzing data collected in complex network environments, closely aligning with numerous real-world applications.

Another closely related line of research is  the high dimensional
vector autoregression (VAR) model \citep{davis2016sparse,
  wang2022high, miao2023high}. By stacking elements in the matrix
$\bY_t = (Y_{ijt})\in \mR^{N_1\times N_2}$ as a vector, the GMNAR
model~\eqref{eq:model} can be re-written as a VAR model with a
dimension of $N_1N_2$. This allows the application of existing techniques for high-dimensional VAR models. While this formulation imposes fewer assumptions on the model coefficients, it leads to a much larger number of parameters that need to be estimated, on the order of $N_1^2N_2^2$, in comparison to $ G(p_1+1)+H(p_2+1)+GH$ (when fixing the group memberships) for the GMNAR model.  Even with commonly used sparsity regularization, the estimation variance of model parameters is likely to be much higher than that of the GMNAR model when the data is generated from the GMNAR model.
For instance, \cite{zhu2022simultaneous} demonstrates that the estimation error based on the regularized sparse VAR model is much larger than that of the Network VAR model when the data is generated from the latter.
Additionally, the coefficients estimated in the general VAR model lack clear interpretations and provide limited insights into network effects. This drawback has also been recognized in existing literature and has motivated the development of matrix-valued autoregressive models~\citep{wang2019factor, chen2021autoregressive,chen2021statistical}.

{A third line of related research is the Spatial Dynamic Panel Data (SDPD) models \citep{yu2008quasi, lee2014efficient,kuersteiner2020dynamic}. For example, \cite{lee2014efficient} explores a model with the formulation $\by_t = \lambda \bW \by_t + \gamma \bW \by_{t-1} + \alpha \by_{t-1} + \bxi^\top \x_{it} + \beps_t$, where $\lambda$ and $\gamma$ represent network effects, $\alpha$ signifies the momentum effect, and $\bxi$ accounts for covariates' impact. These models primarily address only spatial network effects without considering the heterogeneity of spatial locations. In contrast, our proposed model aims to jointly model multiple network effects while accommodating the heterogeneity of network nodes. It is noteworthy that, unlike SDPD models, our framework does not include the cross-sectional spatial network effect (captured by $\lambda$). This aspect could be a promising avenue for future research. Additionally, extending our framework to incorporate time-varying networks and common shocks, as explored in  \cite{kuersteiner2020dynamic}, would also be an interesting research topic.}

Lastly, the proposed GMNAR model has strong connections to recent research in matrix-valued autoregressive models~\citep{wang2019factor, chen2019constrained, chen2021autoregressive, chen2021statistical}. For instance, \cite{chen2021autoregressive} introduced a matrix autoregression (MAR) model to model $\Y_t$, which uses a bilinear autoregressive model of the form $\bY_t = \bA \bY_{t-1} \bB^\top + \bE_t$. Here, $\bA$ and $\bB$ represent unknown autoregressive coefficient matrices, and $\bE_t$ is the noise matrix. In contrast, the proposed GMNAR model can be expressed in an additive form as $\bY_t = \C\Y_{t-1} + \Y_{t-1}\D^\top + \A\circ \Y_{t-1} + \bE_t$, where $\C = \L \bW_1$, $\D = \G \bW_2^\top$, and $\A$ is defined in~\eqref{eq:model}. This highlights that the GMNAR model differs from the matrix-valued autoregressive model of~\cite{chen2021autoregressive}. Importantly, the GMNAR model further explores the (heterogeneous) network structures present in both the rows and columns of the response variable matrix, resulting in a highly interpretable model as illustrated in Model~\eqref{eq:model0}. While the matrix-valued autoregressive model is more interpretable than the generic VAR model, the interpretation of the coefficient matrices can still be less straightforward for many empirical researchers. In this regard, the proposed GMNAR model may be of greater interest to a broader audience across various research disciplines.

}

\section{Model Estimation}\label{sec:estimate}

In this section, we discuss the estimation of the GMNAR model (\ref{eq:model}).
For $g\in [G], h\in [H]$, let
$\btheta_{g}^r = (\lambda_g,
\bzeta_g^\top)^\top\in \mR^{p_1+1}$, $\btheta_h^c = (\gamma_h, \bdelta_h^\top)^\top\in \mR^{p_2+1}$.
Besides, let
$\balpha = (\alpha_{gh})_{g\in [G], h\in  [H]}\in \mR^{G\times H}$
and
$\btheta = (\btheta^{r\top}, \btheta^{c\top},
\vec(\balpha)^\top)^\top \in \mR^{p_1+p_2+GH+2}$.
We also write $\bzeta = (\bzeta_1,\cdots, \bzeta_G)\in \mR^{p_1\times G}$,
$\bdelta = (\bdelta_1,\cdots, \bdelta_H)\in \mR^{p_2\times H}$.
To estimate the model parameters and the group memberships, we aim to minimize the following least squares objective function:
\begin{align}\label{eq:Q_obj}
&Q(\btheta,\mG,\mH) = \sum_{i=1}^{N_1}\sum_{j=1}^{N_2}\sum_{t=1}^T\Big(Y_{ijt} - \lambda_{g_i}\sum_{k = 1}^{N_1} w_{1ik}Y_{kj(t-1)}
-
\gamma_{h_j}\sum_{k = 1}^{N_2} Y_{ik(t-1)}w_{2kj}\nonumber\\
&~~~~~~~~~~~~~~~~~~~~~~~~~~~~~~~~~~~- \alpha_{g_ih_j}Y_{ij(t-1)}
-\bx_{it}\trans\bzeta_{g_i}-\bz_{jt}\trans\bdelta_{h_j}\Big)^2.
\end{align}
We first discuss the estimation when the group memberships
($\mG$ and $\mH$) are given.
The minimization~\eqref{eq:Q_obj} yields that
\bse
\frac{\partial Q(\btheta, \mG,\mH)}{\partial \btheta_g^r} = \zero,~~
\frac{\partial Q(\btheta, \mG,\mH)}{\partial \btheta_h^c} = \zero,~~
\frac{\partial Q(\btheta, \mG,\mH)}{\partial \alpha_{gh}} = 0
\ese
for all $g\in[G], h\in[H]$.
Let
\begin{align}
&\bX_t = (\bx_{1t},\bx_{2t},\cdots, \bx_{N_1t})^\top\in \mR^{N_1\times p_1},\nonumber\\
& \bZ_t = (\bz_{1t},\bz_{2t},\cdots, \bz_{N_2t})^\top\in \mR^{N_2\times p_2},\nonumber\\
& \mX_{ght} =
\Big(\vec(\bW_1^{(\cR_g,\cdot)}\bY_{t-1}^{(\cdot, \mC_h)}),
\one_{N_{2h}}\otimes \bX_{t}^{(\cR_g,\cdot)}\Big)\in \mR^{(N_{1g}N_{2h})\times (p_1+1)},\nonumber\\
& \mZ_{ght} =
\Big(\vec(\bY_{t-1}^{(\cR_g, \cdot)}\bW_2^{(\cdot,\mC_h)}),
\bZ_{t}^{(\mC_h,\cdot)}\otimes\one_{N_{1g}} \Big)\in \mR^{(N_{1g}N_{2h})\times (p_2+1)},\label{eq:tmp}
\end{align}
where $N_{1g} = |\cR_g|$ and $N_{2h} = |\mC_h|$ are defined in the notations. Then one can verify that
\begin{align}
&\frac{\partial Q(\btheta, \mG,\mH)}{\partial \btheta_{g}^r} =
\Big(\sum_{t, h} \mX_{ght}^\top \mX_{ght}\Big)
\btheta_{g}^r
-\sum_{t, h}\mX_{ght}^\top \Big(
\mY_{ght} - \mY_{gh(t-1)}\alpha_{gh} - \mZ_{ght}\btheta_h^c\Big),\label{eq:theta_g_grad}\\
&\frac{\partial Q(\btheta, \mG,\mH)}{\partial \btheta_{h}^c} =
\Big(\sum_{t, g} \mZ_{ght}^\top \mZ_{ght}\Big)
\btheta_{h}^c
-\sum_{t, g}\mZ_{ght}^\top \Big(
\mY_{ght} - \mY_{gh(t-1)}\alpha_{gh} - \mX_{ght}\btheta_g^r\Big),
\label{eq:theta_h_grad}
\end{align}
where $\mY_{ght} = \vec(\bY_{t}^{(\cR_g, \mC_h)})\in \mR^{|\cR_g||\mC_h|}$.
Furthermore, it holds that,
\begin{align}
&\frac{\partial Q(\btheta, \mG,\mH)}{\partial \alpha_{gh}} =
\sum_t \|\mY_{gh(t-1)}\|^2\alpha_{gh}
-\sum_t\mY_{gh(t-1)}^\top \Big(
\mY_{ght} - \mX_{ght}\btheta_g^r - \mZ_{ght}\btheta_h^c\Big).\label{eq:alpha_grad}
\end{align}
{Equations \eqref{eq:theta_g_grad}--\eqref{eq:alpha_grad} define a system of linear equations, whose solutions has the form $\wh \btheta = \bM^{-1}\b$, where
	\begin{align}
		\bM =  \left(
		\begin{array}{ccc}
			\bM^r& \bM^{rc} & \bM^{r\alpha} \\
			\bM^{rc\top}& \bM^c& \bM^{c\alpha}\\
			\bM^{r\alpha\top}& \bM^{c\alpha\top}& \bM^{\alpha}
		\end{array}
		\right),~~~
		\b =  \left(
		\begin{array}{c}
			\b^r \\
			\b^c\\
			\b^{\alpha}
		\end{array}
		\right), \text{ with }\label{eq:Mb}
	\end{align}
the detailed expression of each term shown in Appendix \ref{sec:estimator_form}.}
Subsequently, given the estimated parameters, we update the
group memberships $\mG$ and $\mH$ iteratively.
First, given  $\btheta$ and $\mH$, the $\mG$ is updated by
\begin{align}
\wh g_i  \in &\arg\min_{g_i\in [G]} \sum_{j=1}^{N_2}\sum_{t=1}^T\Big\{Y_{ijt} - \lambda_{g_i}\sum_{k = 1}^{N_1} w_{1ik}Y_{kj(t-1)}
-
\gamma_{h_j}\sum_{k = 1}^{N_2} Y_{ik(t-1)}w_{2kj}
\nonumber\\
&~~~~~~~~~~~~~~~~~~~~~~~~~~~~~~~~~~~- \alpha_{g_ih_j}Y_{ij(t-1)}
-\bx_{it}\trans\bzeta_{g_i}-\bz_{jt}\trans\bdelta_{h_j}\Big\}^2
\label{eq:g_i}
\end{align}
for $i \in [N_1]$.
It's noticeable that the update equation (\ref{eq:g_i}) only involves the row subject $i$ and does not depend on other row subjects. Therefore, it can be executed in a computationally efficient manner.
Similarly, given $\btheta$ and $\mG$, we update $\mH$ by
\begin{align}
\wh h_j  \in &\arg\min_{h_j\in [H]} \sum_{i=1}^{N_1}\sum_{t=1}^T\Big\{Y_{ijt} - \lambda_{g_i}\sum_{k = 1}^{N_1} w_{1ik}Y_{kj(t-1)}
-
\gamma_{h_j}\sum_{k = 1}^{N_2} Y_{ik(t-1)}w_{2kj}\nonumber\\
&~~~~~~~~~~~~~~~~~~~~~~~~~~~~~~~~~~~- \alpha_{g_ih_j}Y_{ij(t-1)}
-\bx_{it}\trans\bzeta_{g_i}-\bz_{jt}\trans\bdelta_{h_j}\Big\}^2.\label{eq:h_j}
\end{align}
for $j \in [N_2]$.
We summarize the algorithm in Algorithm \ref{alg:gmnar}. The algorithm comprises iterations of two major steps. The first step involves estimating the parameters with given group memberships, while the second step focuses on updating group memberships given the parameters. Each step can be computed efficiently due to the simple analytical forms.
The algorithm can be validated to converge to a local minimizer, of which the proof is given in Appendix \ref{sec:alg_converge}.
Additionally, we discuss how to obtain the initial estimators in {Algorithm \ref{alg:init}}. It's worth noting that the algorithm requires the group numbers $G$ and $H$ to be specified initially. Therefore, we provide a criterion to estimate the true group numbers, $G_0$ and $H_0$, and establish the selection consistency thereafter.

\subsection{Selection of Group Numbers}\label{sec:group_number}

We now discuss the estimation of group numbers
$G$ and $H$.
To simplify the notations, we write $\wh \btheta^{(G,H)}, \wh \mG^{(G,H)}, \wh \mH^{(G,H)}$ as the estimators when the row and column group numbers are specified as $G$ and $H$, respectively. Then, we estimate $G_0$ and $H_0$ by utilizing the following information criterion:
\beq
\qic(G, H) = \log\{Q(\wh \btheta^{(G,H)}, \wh \mG^{(G,H)}, \wh \mH^{(G,H)})\} + \lambda(G,H),\label{eq:qic}
\eeq
where $\lambda(G,H)$ is a penalty function.
Then we estimate the group numbers by
$(\wh G, \wh H)  \in \arg\min_{(G,H)} \qic(G,H)$.
In practice, we specify $\lambda(G,H) = \kappa(G+H)$ and
in the theoretical analysis, we show that as long as
$T^{-1/2}m\ll \kappa \ll c_\gap c_\pi^2/(GH)$, we can estimate
$G_0$ and $H_0$ consistently with the QIC criterion.
Here $c_\gap, c_\pi$ are group related values related to model signals,
 which will be introduced later in our theoretical analysis.
In our numerical study, we specify $\kappa = \{C (\log T) T^{1/8}\}^{-1}$ with $C = 40$, which is able to achieve a reliable finite sample performance.
 In addition, to check the robustness of our penalty function specifications, we follow
 \cite{liu2020identification} for verifying different $C$ values.
 The results are included in Appendix \ref{subsec:simu:tuning} and it shows that the true group numbers can still be estimated with high accuracy when the sample sizes are large.
It would be interesting to develop a tuning free procedure for the group number selection, which we leave as an interesting future topic.

\begin{algorithm}
	\caption{Estimation of the GMNAR Model}
	\label{alg:gmnar}
	\begin{algorithmic}[1]
		\State {\bf Input:} $\{\bY_t, \bX_t, \bZ_t\}$, $\{\bW_1, \bW_2\}$, and $\{G, H\}$.
		\State Obtain initial group memberships $\mG^{(0)}$ and $\mH^{(0)}$ according to Algorithm \ref{alg:init}. Let $\{\btheta^{(k)}, \mG^{(k)}, \mH^{(k)}\}$ be the estimators and memberships in the $k$th iteration.
 \State     Repeat {\sc Step 1} and {\sc Step 2} for {$k = 1, \cdots$}
 until convergence.

 {\sc Step 1.} Given $\{\mG^{(k-1)}, \mH^{(k-1)}\}$, calculate
       $ \btheta^{(k-1)} = (\bM^{(k-1)})^{-1} \b^{(k-1)}$, where $\bM^{(k-1)}$ and $\b^{(k-1)}$ are obtained from
       (\ref{eq:Mb}) with $\{\mG^{(k-1)}, \mH^{(k-1)}\}$ specified.

{\sc Step 2.} Given $ \btheta^{(k-1)}$, update the memberships by (\ref{eq:g_i}) and (\ref{eq:h_j}) to obtain $\{\mG^{(k)}, \mH^{(k)}\}$.

 	    \State {\bf Output:} Final estimator and memberships:
$\wh\btheta = \btheta^{(K)}$,
$\wh \mG = \mG^{(K)}$,
$\wh \mH = \mH^{(K)}$.
Here $K$ is the final number of iteration rounds.
	\end{algorithmic}
\end{algorithm}

\begin{algorithm}
\caption{Initialization of the GMNAR Model}
	\label{alg:init}
	\begin{algorithmic}[1]
		\State {\bf Input:} $\{\bY_t, \bX_t, \bZ_t\}$, $\{\bW_1, \bW_2\}$, and $\{G, H\}$.
		\State Treat each node as a group, estimate $\wh\btheta^r, \wh\btheta^c$ and $\wh\balpha$ by \eqref{eq:Mb}.
		\State  Run the $k$-means clustering for the above estimators for $T_{\text{init}}=3$ trials. For each trial $t = 1, \cdots, T_{\text{init}}$, try the following two clustering types:
		
		{\bf Type 1. (Clustering by time effect)}
		
		{\sc Step 1.} Clustering the first type of nodes by the
		{\it self-driven time effect} $\wh\balpha=
		(\wh\balpha_{1 \cdot}^{ \top}, \cdots,
		\wh\balpha_{N_1 \cdot}^{ \top})^\top$; clustering the
		second type by $\wh\balpha^{\top} =
		(\wh\balpha_{\cdot 1},\cdots, \wh\balpha_{\cdot N_2})^\top$,
		where $\wh\balpha_{i \cdot}^{ \top}$ is the $i$th row of
		$\wh\balpha$, and  $\wh\balpha_{\cdot j}^{ \top}$ is the $j$th column
		of $\wh\balpha$.
		
		{\sc Step 2.} Calculate the loss in the $t$th trial for type 1 by $Q^{(t)}(\wh\btheta, \mG^{(t)}_1, \mH^{(t)}_1)$.
		
		{\bf Type 2. (Clustering by network effect)}
		
		{\sc Step 1.} Clustering by {\it row(column) network
			effects} on  $\wh\btheta^{r} = (\wh\btheta^r_1,\cdots,
		\wh\btheta^r_{N_1})$ and {\it covariate effects} on $\wh\btheta^{c} =
		(\wh\btheta^c_1,\cdots, \wh\btheta^c_{N_2})$.
		
		{\sc Step 2.} Calculate the loss in the $t$th trial for type 2 by $Q^{(t)}(\wh\btheta, \mG^{(t)}_2, \mH^{(t)}_2)$.

		\State Select the best initial trial $t^* = \argmin_{t} [\min\{ Q^{(t)}(\wh\btheta, \mG^{(t)}_1, \mH^{(t)}_1), Q^{(t)}(\wh\btheta, \mG^{(t)}_2, \mH^{(t)}_2)\}$], and the corresponding initial memberships are denoted as $\mG^{(0)}, \mH^{(0)}$.
		\State {\bf Output:} Best initial memberships: $\mG^{(0)}, \mH^{(0)}$.
	\end{algorithmic}
\end{algorithm}

\section{Theoretical Properties}\label{sec:theory}

\subsection{Estimation Consistency}

Define $\bTheta_{ij} = (\btheta_{g_i}^{r\top}, \btheta_{h_j}^{c\top}, \alpha_{g_ih_j})^\top\in \mR^{p_1+p_2+3}$ and
$\bTheta = (\bTheta_{ij}:i \in [N_1], j\in [N_2])$
as a tensor of dimension $N_1\times N_2\times (p_1+p_2+3)$.
With $\bTheta$ we can rewrite (\ref{eq:Q_obj}) as
\beq
Q(\bTheta) = \sum_{i  =1}^{N_1}\sum_{j = 1}^{N_2} \sum_{t = 1}^T
\big(Y_{ijt} - \cX_{ijt}^\top\bTheta_{ij}\big)^2\defeq
\sum_{i  =1}^{N_1}\sum_{j = 1}^{N_2} Q_{ij}(\bTheta_{ij})
.\label{def:Qij}
\eeq
Denote by $\wh \bTheta = (\wh \bTheta_{ij} =
(\wh \btheta_{\wh g_i}^{r\top}, \wh \btheta_{\wh h_j}^{c\top}, \wh \alpha_{\wh g_i\wh h_j})^\top)$ as the global minimizer of $Q(\bTheta) $, we define
the following pseudo distance as
\begin{align}
d(\wh \bTheta, \bTheta) &= \frac{1}{N_1N_2} \sum_{i = 1}^{N_1}
\sum_{j = 1}^{N_2} \Big\|\wh \bTheta_{ij} - \bTheta_{ij}\Big\|^2\nonumber\\
& = \frac{1}{N_1}\sum_{i = 1}^{N_1}  \|\wh \btheta_{\wh g_i}^{r} - \btheta_{g_i}^{r} \|^2 +
\frac{1}{N_2}\sum_{j = 1}^{N_2} \|\wh \btheta_{\wh h_j}^c - \btheta_{h_j}^c\|^2 + \frac{1}{N_1N_2}\sum_{i,j} |\wh \alpha_{\wh g_i\wh h_j} - \alpha_{g_ih_j}|^2.\label{eq:pseudo_dist}
\end{align}
Therefore the $d(\wh \bTheta, \bTheta)$
measures the average distance between $\wh\bTheta$ and $\bTheta$.
In the following we first establish the consistency of $\wh \bTheta$ using this pseudo metric.
To this end, we require the following definition and conditions.

\begin{definition}\label{def:convex_concen}
	($K$-Convex concentration property)
	Let $\bx\in \mR^n$ be a random vector.
	If for every 1-Lipschitz convex function $\varphi: \mR^n\to \mR$, we have
	$E|\varphi(\bx)|<\infty$ and for every $t>0$
	\beq
	P\Big(\Big|\varphi(\bx) - E\{\varphi(\bx)\}\Big|\ge t\Big)\le 2\exp(-t^2/K^2),\nonumber
	\eeq
	then $\bx$ is named to have the $K$-convex concentration property.
\end{definition}


\begin{assumption}\label{assum:para_space}
 {\sc (Parameter Space)} The parameter satisfies that $\|\bTheta\|_{\max} < \infty$.
\end{assumption}

\begin{assumption}\label{assum:tau_min}
{\sc(Convexity)}
Let $ \cX_{ijt} \defeq
(\sum_{k = 1}^{N_1} w_{1ik}Y_{kj(t-1)}, \bx_{it}^\top,
\sum_{k = 1}^{N_2} Y_{ik(t-1)}w_{2kj}, \bz_{jt}^\top,\\
Y_{ij(t-1)})^\top\in \mR^{p_1+p_2+3}$ and
  let $\bSigma_{ij} = E(\cX_{ijt}\cX_{ijt}^\top)$ and
$\tau_{\min}\defeq \min_{i,j}\lambda_{\min}(\bSigma_{ij})$
is a positive constant.
\end{assumption}

\begin{assumption}\label{assum:sub_gaussian}
{\sc(Distribution of Noise Term)}
 Assume $\ve_{ijt}$ is {\it i.i.d} across $i\in [N_1]$, $j\in [N_2]$,
 and $t\in [T]$. In addition, $\ve_{ijt}$ is a zero-mean sub-Gaussian
 variable with a scale factor $0<\sigma<\infty$, i.e.,
 $E\{\exp(u\ve_{ijt})\}\le \exp(\sigma^2u^2/2)$.
Let $\ve_{ijt}$ be independent of $
\{\bY_s: s\le t-1\}$, $\{\bX_s = (\bx_{is}: i\in [N_1])^\top: s\le t\}$, and
$\{\bZ_s = (\bz_{js}: j\in [N_2])^\top: s\le t\}$.
\end{assumption}

\begin{assumption}\label{assum:mixing}
{\sc(Distribution of Covariates)}
 Assume $E(\bx_{it}) = \zero$ and $E(\bz_{jt}) = \zero$
  for any $i\in [N_1]$, $j\in [N_2]$ and $t \in [T]$.
  Let $\bfeta_1 \in \mR^{p_1}$ (and $\bfeta_2 \in \mR^{p_2}$) be a constant vector satisfying $\|\bfeta_1\| = 1$ ($\|\bfeta_2\| = 1$).
Define $\bx_t^\eta = (\bx_{it}^\top\bfeta_{1}: i\in [N_1])^\top \in \mR^{N_1}$,
$\bz_t^\eta = (\bz_{jt}^\top\bfeta_{2}:j\in [N_2])^\top \in \mR^{N_2}$.
 Assume $\{(\bx_t^{\eta\top}, \bz_t^{\eta\top})^\top: 0\le t\le T\}^\top$
satisfies the $K$-convex concentration property for some constant $K$
according to Definition \ref{def:convex_concen}.
\end{assumption}

\begin{assumption}\label{assum:station}
{\sc(Stability)}
  Assume $\mY_0 = \zero$, and assume that $\max_{g\in [G_0],h\in [H_0]} |\lambda_g^0 + \gamma_h^0 + \alpha_{gh}^0|\le c_{\max} <1$,
  where $G_0$ and $H_0$ are true number of groups and $c_{\max}$ is a positive constant.
\end{assumption}

\begin{assumption}\label{assum:group_diff}
{\sc(Group Difference)}
  $\min_{g_1\ne g_2}\{\|\btheta_{g_1}^{r0}  - \btheta_{g_2}^{r0}\|^2 +
   \max_{h\in [H_0]} |\alpha_{g_1h}^0 - \alpha_{g_2h}^0 |^2\}\ge  c_\gap$
  and $\min_{h_1\ne h_2}\{\|\btheta_{h_1}^{c0}  - \btheta_{h_2}^{c0}\|^2 +\max_{g\in [G_0]}|\alpha_{gh_1}^0 - \alpha_{gh_2}^0 |^2\}\ge c_\gap$,
  where $c_\gap>0$ and is allowed to go to zero as $N_1, N_2\to \infty$.
\end{assumption}


\begin{assumption}\label{assum:group_ratio}
{\sc(Group Proportion)}
Let $\{g_i^0: i\in [N_1]\}$ and $\{h_j^0: j\in [N_2]\}$ be non-random true membership sequences.
Let $\pi_{g, N_1}^r = \sum_i I(g_i^0 = g)/N_1$ and $\pi_{h, N_2}^c = \sum_j I(h_j^0 = h)/N_2$ for $g\in [G_0]$
and $h\in [H_0]$.
Assume that $\min_{g \in [G_0], h \in [H_0]} \{\pi_{g, N_1}^r, \pi_{h,
	N_2}^c\} \ge c_{\pi} > 0$ for sufficiently large $N_1, N_2$.
Here $c_{\pi}$ is allowed to go to zero as $\min\{N_1, N_2\}\to \infty$.
\end{assumption}

The Assumption \ref{assum:para_space} requires the parameter space to be bounded.
Assumption \ref{assum:tau_min} ensures the convexity of the pairwise objective function, i.e.,
$Q_{ij}(\bTheta_{ij})$, as a function of $\bTheta_{ij}$ for sufficiently
  large $T$.
This condition is crucial for establishing the consistency result for the pseudo distance in \eqref{eq:pseudo_dist}.

 Assumptions \ref{assum:sub_gaussian}--\ref{assum:mixing} concern about the distribution of the noise term and covariates respectively.
Specifically, Assumption \ref{assum:sub_gaussian} requires the noise term $\ve_{ijt}$ to be {\it i.i.d.} and it follows sub-Gaussian distribution, which is widely used in high dimensional time series literature \citep{wang2013calibrating,lugosi2019sub,fan2021augmented}.
We also provide a weighted least squares estimation procedure
with group-specific variances, i.e., $\var(\ve_{ijt}) = \sigma_{g_i^0h_j^0}^2$ in Appendix \ref{sec:wls}.
Subsequently, Assumption \ref{assum:mixing} allows the covariates $\{\bx_{it}\}$ and $\{\bz_{jt}\}$ to be correlated but satisfying the  $K$-convex concentration property according to Definition \ref{def:convex_concen}.
This assumption is employed to establish Hanson-Wright type inequality for dependent variables \citep{adamczak2015note}.
Although this is a high level condition, there are a variety of random variables satisfying Definition \ref{def:convex_concen}, as discussed in the following Remark \ref{remark.K_convex}.
We further comment that the Assumptions \ref{assum:sub_gaussian}--\ref{assum:mixing} together imply that
$ \bv^\eta \defeq (\bv_t^\eta: 0\le t\le T)^\top$
  satisfies the $K$-convex concentration property for some constant $K$.
Here $ \bv_t^\eta = (\bx_t^{\eta\top}, \bz_t^{\eta\top},
  \E_t^\top)^\top$, where $\E_t = \vec(\bE_t)$.

\begin{remark}\label{remark.K_convex}
As discussed by \cite{adamczak2015note},
there are a variety of random vectors $\bx$ satisfying the $K$-convex concentration property in Definition \ref{def:convex_concen}. For example, (i) Any random vector $\bx$ with its elements $x_i$s independent for all $i$, and $|x_i| \le 1 ~ a.s.$ satisfies Definition \ref{def:convex_concen} \citep{talagrand1988isoperimetric}; (ii) Any random vector $\bx$ with its elements in a bounded interval and geometrically strongly mixing satisfies Definition \ref{def:convex_concen} \citep{samson2000concentration}.
We refer to \cite{adamczak2015note} for more detailed discussions.
\end{remark}

Next, Assumption \ref{assum:station} ensures the stability of the
matrix-valued time series data as $T$ goes to infinity, as
defined in \cite{lutkepohl2005new}.
Assumptions \ref{assum:group_diff} and \ref{assum:group_ratio} are imposed on certain group properties. Condition \ref{assum:group_diff} assumes there is a gap between true parameters of two different groups.
The condition is an extension of the same type condition assumed by the group panel data models with only one group specified \citep{su2016identifying,ando2016panel,zhang2019quantile,liu2020identification}.
The special care is paid for the autoregression parameter $\alpha_{gh}$, where the row and column groups are both involved.
Therefore we require a min-max type condition for $\alpha_{gh}^0$ in Assumption \ref{assum:group_diff}.
Furthermore, instead of assuming $c_{\gap}>c>0$ by a positive constant $c$ in existing literature \citep{zhang2019quantile,liu2020identification},
we allow $c_{\gap}\to 0$ to study how this signal strength affects the theoretical properties.
Lastly, Assumption \ref{assum:group_ratio} assumes that there is a lower bound of row and column group proportions.
Here we allow $c_{\pi}\to 0$, which indicates that we may have diverging group numbers with $G_0,H_0\to \infty$.
In the following theoretical analysis, we further specify a lower bound of the convergence speed of $c_\pi$ to ensure the estimation consistency result.
This is also a relaxed condition than existing literature where a fixed number of groups is typically assumed \citep{su2016identifying,liu2020identification}.
In the following we establish the consistency result for the pseudo distance.

\bet\label{thm:pseudo_dist}
Suppose $G\ge G_0$ and $H\ge H_0$, where $G_0$ and $H_0$ are
true number of groups.
In addition, assume Assumptions \ref{assum:para_space}--\ref{assum:station} hold.
Then it follows,
\beq
d(\wh\bTheta, \bTheta^0)  = O_p(T^{-1/2}(m+\log (N_1N_2))),\nonumber
\eeq
where $\bTheta^0  = (\bTheta_{ij}^0 =
(\btheta_{g_i^0}^{r0\top},
\btheta_{h_j^0}^{c0\top},
\alpha_{g_i^0h_j^0}^0)^{\top})$ and $m=p_1 + p_2 +3$.
\eet

Theorem \ref{thm:pseudo_dist} implies that as long as we have
$T^{1/2}\gg \log(N_1N_2)+m$, $\wh\bTheta$ is a consistent estimator
for $\bTheta^0$ in the metric of pseudo distance when $G$ and $H$ are sufficiently large.
It is remarkable that the consistency result holds when the group
numbers $G$ and $H$ are larger than or equal to $G_0$ and $H_0$.
Furthermore, we discuss in Theorem \ref{thm:select_GH} that the QIC can
consistently select the true group numbers when the tuning
parameters are properly specified.


\bet\label{thm:select_GH}
Assume Conditions \ref{assum:para_space}--\ref{assum:group_ratio}
hold and let $c_\gap c_\pi^2\gg T^{-1/2}(m+\log (N_1N_2H))GH$. In addition, assume $\kappa = \lambda(G,H)/(G+H)$ satisfies
\beq
T^{-1/2}(m+\log (N_1N_2H))\ll \kappa \ll c_\gap c_\pi^2/(GH). \label{eq:eta_range}
\eeq
Then we have $P(\wh G = G_0, \wh H = H_0)\to 1$ as $\min(N_1,N_2,T)\to \infty$.
\eet

Theorem \ref{thm:select_GH} implies that if we set $\kappa$ to satisfy (\ref{eq:eta_range}), then we can consistently estimate the true group numbers. Specifically, we need $\kappa\gg T^{-1/2}(m+\log (N_1N_2H))$ to ensure that $\qic(G, H)>\qic(G_0,H_0)$ when $G<G_0$ or $H<H_0$. Conversely, we need $\kappa\ll c_\gap c_\pi^2/(GH)$ to guarantee that $\qic(G, H)>\qic(G_0,H_0)$ when $G>G_0$ and $H>H_0$.
We further remark that \eqref{eq:eta_range} explicitly requires that
$c_\pi^2\gg T^{-1/2}(m+\log (N_1N_2H))GHc_\gap^{-1}$ for the estimation consistency result, which specifies a lower bound requirement for
$c_\pi$ to converge to zero.
When both conditions are met, we can obtain $\wh G = G_0$ and $\wh H = H_0$ with a probability approaching one with a large sample size. In the next subsection, we further discuss the results of node-wise parameter estimation and the consistency of group membership estimation.

\subsection{Group Membership Estimation Consistency}\label{sec:42}

As we stated before, the pseudo distance in (\ref{eq:pseudo_dist})
measures the average distance between $\wh \bTheta$ and $\bTheta^0$.
Therefore, the result in Theorem \ref{thm:pseudo_dist} is not
sufficient to imply the parameter consistency for each node.
To this end, we derive the following node-wise parameter consistency
result, which will be crucial to build the membership estimation
consistency later.

\bep\label{pro:gh_consistency}
Assume Assumptions \ref{assum:para_space}--\ref{assum:station} hold.
When $G\ge G_0$ and $H\ge H_0$, we have
\begin{align}
& \sup_j\left\{\|\wh \btheta_{\wh h_j}^c - \btheta_{h_j^0}^{c0}\|^2
 +\frac{1}{N_1}\sum_{i} |\wh \alpha_{\wh g_i\wh h_j} - \alpha_{g_i^0h_j^0}^0|^2\right\} = O_p(c_\pi^{-1}T^{-1/2}(m+\log (N_1N_2))),\label{eq:sup_j_para_diff}\\
 & \sup_i\left\{\|\wh \btheta_{\wh g_i}^r - \btheta_{g_i^0}^{r0}\|^2
 +\frac{1}{N_2}\sum_{j} |\wh \alpha_{\wh g_i\wh h_j} - \alpha_{g_i^0h_j^0}^0|^2\right\} = O_p(c_\pi^{-1}T^{-1/2}(m+\log (N_1N_2))).\label{eq:sup_i_para_diff}
\end{align}
\eep
It's worth noting that equations (\ref{eq:sup_j_para_diff}) to (\ref{eq:sup_i_para_diff}) establish uniform node-wise parameter estimation consistency, which is crucial for achieving group membership consistency for $\wh \mG$ and $\wh \mH$ when $G\ge G_0$ and $H\ge H_0$. Since the discussions are similar for both $\wh \mG$ and $\wh \mH$, we will illustrate using $\wh \mH$ in the following. Given $H_0, G_0, \mH^0, \mG^0, \btheta_0$, and $H, G, \mH, \mG, \btheta$, we first define the following pseudo distance as a measure of dissimilarity between two sets:
\begin{align}
d_S(\btheta, \btheta^0; \mG, \mG^0) =
\max\Big\{&\max_{h_0\in [H_0]}\min_{h \in [H]}
\Big(\|\btheta_h^c - \btheta_{h_0}^{c0} \|^2+
\frac{1}{N_1}\sum_i|\alpha_{g_ih} -
\alpha_{g_i^0 h_0}^0|^2\Big),\nonumber\\
&\max_{h \in [H]}\min_{h_0\in [H_0]}
\Big(\|\btheta_h^c - \btheta_{h_0}^{c0} \|^2+
\frac{1}{N_1}\sum_i|\alpha_{g_ih} -
\alpha_{g_i^0 h_0}^0|^2\Big)\Big\}.\label{eq:distance_set}
\end{align}
Define an $\eta$-neighbourhood for $\btheta^0$ based on the
above distance as $\mN_\eta = \{\btheta:
d_S(\btheta, \btheta^0; \mG, \mG^0)<\eta\}$.
For $\btheta\in \mN_\eta$, denote the sets
\bse
\mA_\eta(\btheta, h_0) = \left\{h\in [H]:
\|\btheta_{h}^{c} - \btheta_{h_0}^{c0} \|^2+
\frac{1}{N_1}\sum_i| \alpha_{\wh g_ih} -
\alpha_{g_i^0 h_0}^0|^2\le \eta\right\},
\ese
for $h_0 \in [H_0]$.
Here $\mA_\eta(\btheta, \cdot)$ is used to map the memberships in $[H_0]$
to $[H]$.
We then establish the membership estimation consistency result as follows.

\bet\label{thm:h_consistency2}
Assume Assumptions \ref{assum:para_space}--\ref{assum:group_ratio}, $G\ge G_0$,
and $H\ge H_0$.
Suppose we have $d(\bTheta, \bTheta^0) = O_p(T^{-1/2}(m+\log (N_1N_2)))$
and $c_\gap c_\pi \gg d(\bTheta, \bTheta^0)$ as $\min(N_1, N_2) \to \infty$.
Then the following conclusions hold:\\
(i) For all $\btheta\in \mN_\eta$ with $\eta<c_\pi c_\gap/4$,
$ \{\mA_\eta(\btheta, h_0), h_0 \in [H_0]\}$ is a partition of $[H]$;\\
(ii)
Define the event $\Omega = \{\wh h_j \in \mA_\eta(\btheta, h_j^0), \forall
j\in [N_2]\}$ for $\btheta\in \mN_\eta$ and
$\eta< \tau_{\min}c_\gap
c_\pi/\{4(\tau_{\min}+\tau_{\max})\}$, where $\wh h_j$ is
  defined by (\ref{eq:h_j}).
Then we have
\bse
P(\Omega^c)\le HN_1N_2\exp\Big(-c_1 T^{1/2}c_\gap c_\pi +c_2m\Big),
\ese
where $c_1, c_2$ are two positive constants.\\
(iii) Define  $\{\wh h_j:  j\in [N_2]\}$ by (\ref{eq:h_j}) when $\wh \btheta$ is specified.
Let $T^{1/2}c_{\gap}c_\pi^2 \gg \log(N_1N_2H) + m$, then we have
$\wh \btheta \in \mN_\eta$ and
for each $\wt h\in [H]$, there exists a $h \in [H_0]$,
such that $\wh \mC_{\wt h}\subset \mC_{h}^{0}$ with probability
tending to 1.
\eet

Some comments on Theorem \ref{thm:h_consistency2} are in order.
For any $\btheta\in \mN_\eta$, $\mA_\eta(\btheta,\cdot)$ defines a map from $[H_0]$ to $[H]$. The conclusion (i) implies that for any $h_1\ne  h_2$, we have
$\mA_\eta(\btheta, h_1)\cap \mA_\eta(\btheta, h_2) = \emptyset$ as long as $\eta$ is sufficiently small.
Next, the conclusion (ii) states that with a high probability, the event
$\Omega$ will hold. Specifically,
for $T^{1/2}c_{\gap}c_\pi \gg \log(N_1N_2H) + m$, we have
$P(\Omega^c)\to 0$.
Subsequently, in conclusion (iii), we require a stronger condition to
ensure that $\wh \btheta\in \mN_\eta$.
As implied by the conclusion (iii), the true groups are split into subgroups
instead of joining into new groups  when $H\ge
  H_0$.
With similar arguments we can show that
for each $\wt g\in [G]$, there exists a $g \in [G_0]$,
such that $\wh \cR_{\wt g}\subset  \cR_{g}^{0}$ with probability
tending to 1 when $G\ge G_0$.

Define $\wh\mG = (\wh g_i: i\in [N_1])^\top \in \mR^{N_1}$
and $\wh\mH = (\wh h_j: j\in [N_2])^\top \in \mR^{N_2}$ as the estimated membership vectors.
Particularly, for $G = G_0$ and $H = H_0$, we can show that
$\wh \mG = \mG^0$ and $ \wh \mH = \mH^0$ hold with probability tending to 1.
Furthermore, let $\wh\btheta^{\o}$ be the oracle estimator when
  the true group memberships $\mG^0$ and $\mH^0$ are known.
  Then the oracle property holds that $\wh \btheta = \wh \btheta^{\o}$ with probability tending to 1. The result is presented in the following Corollary.

\begin{corollary}\label{coro:group_consistency}
  Assume Assumptions \ref{assum:para_space}--\ref{assum:group_ratio},
  and $G = G_0$ $H = H_0$.
Assume $T^{1/2}c_\gap c_\pi^2\gg\log(N_1N_2H)+m$.
Then under label permutations, we have
  \begin{align}
  &\lim_{\min(N_1,N_2,T)\to \infty} P\left(\wh \mG = \mG^0, \wh \mH = \mH^0
  \right)\to 1, \label{eq:membership_consistency}\\
  &  \lim_{\min(N_1,N_2,T)\to \infty}  P\left(\wh \btheta = \wh \btheta^{\o}
    \right)\to 1. \label{eq:oracle}
  \end{align}
\end{corollary}

The results in Corollary \ref{coro:group_consistency} imply that $\wh \btheta$ is asymptotically equivalent to $\wh \btheta^\o$. Therefore, to derive the asymptotic distribution of $\wh\btheta$, it is sufficient to investigate
The details are given in the subsequent section.

\subsection{Asymptotic Normality}

Next, we turn our attention to the statistical inference of model parameters.
To facilitate this discussion, we will assume the following condition.
\begin{assumption}\label{assum:n}
  Assume there exists $n$ so that
\beq
c_1n\le \min_{g,h}\{\min(N_{1g}, N_{2h})\}\le \max_{g,h} \{\max(N_{1g}, N_{2h})\}\le
c_2n,\nonumber
\eeq
where $c_1,c_2>0$ are constants,  and $n\to \infty$ when $N_1, N_2\to\infty$.
\end{assumption}
In other words,  we assume that all $N_{1g}$ ($g\in[G]$) and
  $N_{2h}$ ($ h\in[H]$) to diverge at the same rate
$n$, so that a balance between group sizes is achieved.
We establish the asymptotic normality of the estimator in
  the following Theorem for subsequent statistical
  inference.

\bet\label{thm:normal}
Assume Assumptions \ref{assum:para_space}--\ref{assum:n},
  and $G = G_0$ $H = H_0$.
Assume $T^{1/2}c_\gap c_\pi^2\gg\log(N_1N_2 \max(H, G))+m$.
Define $\bLambda = \diag\{(nN_2T)^{-1/2}\I_{G(1+p_1)},
(nN_1T)^{-1/2}\I_{H(1+p_2)},n^{-1}T^{-1/2}\I_{GH}\}$,
$\bM_{\Lambda, NT}^0 = \bLambda E(\bM)\bLambda$ and assume $ \bM_{\Lambda}^0 = \lim_{\min(N_1,N_2,T)\to \infty} \bM_{\Lambda, NT}^0$ exists.
Assume $\lambda_{\min}(\bM_\Lambda^0)\ge \tau>0$ and $q^{3/2}/\sqrt
T\to 0$,
where $q = G(1+p_1)+H(1+p_2)+GH$.
Then for any $\bfeta\in \mR^{q}$ with $\|\bfeta\| = 1$
we have
\beq
\bfeta^\top\bLambda^{-1}(\wh\btheta - \btheta^0) \to_d N\left(\zero, \sigma^2\bfeta^\top  (\bM_\Lambda^0)^{-1}\bfeta\right).\label{eq:normal0}
\eeq
\eet

Theorem \ref{thm:normal} establishes the asymptotic normality of the estimator.
Specifically, the convergence rates of $\wh\btheta^r$ and $\wh \btheta^c$ are
$\sqrt{nN_2T}$ and $\sqrt{nN_1T}$ respectively.
They are both faster than $\wh \alpha_{gh}$, which is $n\sqrt
T$-consistent according to (\ref{eq:normal0}).
The difference is due to their different effective sample sizes.
Using \eqref{eq:normal0}, we are able to conduct the statistical inference.

We next provide an estimator to the asymptotic covariance.
With the parameter estimator $\wh\bTheta = (\wh\bTheta_{ij} = (\wh\btheta_{\wh g_i}^{r \top}, \wh\btheta_{\wh h_j}^{c \top}, \wh\alpha_{\wh g_i \wh h_j})^\top)$, we first estimate $\sigma^2$ as follows
\begin{align}
	\wh\sigma^2 = \frac{1}{N_1 N_2 T} \sum_{i=1} \sum_{j = 1} \sum_{t = 1} (Y_{ijt} - \cX_{ijt}^\top \wh\bTheta_{ij})^2, \label{eq:sigma_hat}
\end{align}
where $\cX_{ijt}$ is defined in Assumption \ref{assum:tau_min}.
Next, we estimate $\M_{\Lambda}^0$ by $\bLambda\wh \M\bLambda$,
where $\wh \M$ is obtained  by plugging estimated parameters $\wh \bTheta$ into the expression in \eqref{eq:Mb}.
%
In the following theorem, we show that the covariance estimator is consistent.
\bet\label{thm:cov_consistent}
Suppose Assumption \ref{assum:para_space}--\ref{assum:station} and Assumption \ref{assum:n} hold, and $G = G_0$ $H = H_0$.
Assume $\{m + \log(N_1 N_2)\} / \sqrt{T} \to 0$, where $m = p_1 + p_2 + 3$, and denote $\wh\bM_{\Lambda} = \bLambda \wh\bM \bLambda$. Then the following holds,
\begin{align*}
  \wh\sigma^2 \to_p \sigma^2,~ \text{and } ~ \wh\sigma^2 \bfeta^\top (\wh\bM_\Lambda)^{-1}\bfeta  \to_p \sigma^2 \bfeta^\top (\bM_\Lambda^0)^{-1} \bfeta,
\end{align*}
where $\bfeta$ is defined in Theorem \ref{thm:normal}.
\eet
Theorem \ref{thm:cov_consistent} indicates that
we can obtain a consistent estimator for the
asymptotic variance by plugging in the estimators $\wh \bTheta$ and the consistent estimator $\wh\sigma^2$.
This assures a valid statistical inference procedure.
We next present a number of simulation studies to examine the finite sample performances of the model estimation and inference procedures.

 \section{Simulation Study}\label{sec:simulation}

\subsection{Model Settings}\label{subsec:model_set}
To illustrate how our proposed method performs with a finite sample size, we conduct several simulation studies in this section. We follow the approach outlined in the existing literature \citep{huang2017spatial,
	ren2021graphical, zhu2022simultaneous}, and examine two
      distinct network structures.

{\bf Example 1.} (Stochastic Block Model, SBM) The first type of
network is the stochastic block model \citep{wang1987stochastic,
	nowicki2001estimation}, in which nodes in the same block (group)
are assigned with higher probability to be connected, while
nodes in different blocks are less likely to be connected.
Following the setting of \cite{nowicki2001estimation}, we
first assign a group label randomly with the probability $1/K$ for
each node, where $K$ is the total number of groups. When the $i$th
and the $j$th node are in the same group, we set $P(a_{ij} = 1) =
20/N$, and otherwise we set $P(a_{ij} = 1) = 2/N$.

{\bf Example 2.} (Power-Law Distribution Network) The second
type of network is generated from a power-law distribution following
\cite{clauset2009power}.
For the $i$th node, its in-degree $d_i = \sum_{j = 1}^N a_{ji}$ is
assumed to be power-law distributed. This coincides with the
``super-star'' effect in real world social networks, which refers to
the phenomenon that only few people have a huge number of followers.
Specifically, we first generate $\wt d_i$ from a discrete
power-law distribution with probability $P(\wt d_i = k)
\propto k^{-2.5}$,
and then we set $d_i = 4 \wt d_i$.
Then, $d_i$ followers of the $i$th node are randomly
selected to construct the adjacency matrix.
As a result, the adjacency matrix generated by the power-law distribution network is not symmetric, which implies a directed network.

In both examples, we consider three different scenarios for $G_0$ and $H_0$.
In each scenario, the node memberships are sampled from the
multinomial distribution with probability $\bpi_1 = \{\pi_g =
G_0^{-1}: g = 1,\cdots, G_0\}$ and $\bpi_2 = \{\pi_h = H_0^{-1}: h =
1, \cdots, H_0\}$.
The dimension of exogenous covariates are set as $p_1 = p_2 = 3$, and
the corresponding true parameters are shown in Table
\ref{table:true_param} in Appendix \ref{subsec:appd_simu}. For all scenarios, the covariates $\bx_{it}$
and $\bz_{jt}$ are generated from multivariate normal distribution
$N(\zero, \bI_{p_1})$ and $N(\zero, \bI_{p_2})$, respectively. Lastly,
we generate the noise term $\ve_{ijt}$ from $N(0,1)$ independently.

\subsection{Performance Measure and Simulation Results}\label{subsec:simu_res}

In this section, we first introduce the model performance measure and
then present the simulation results.
We set the network sizes  $(N_1, N_2) \in \{(100, 80),(200, 150),(300, 250)\}$.
In addition, the time length is set to be $T \in \{20,40\}$.
For each scenario, we repeat the experiments for $R=500$ times.
The networks are fixed throughout all replicates under one setting.
 In the initialization, we use 3 trials for each type of clustering types as described in Algorithm \ref{alg:init}.
Denote the estimated parameters in the $r$th replicate as
$\wh\lambda_g^{(r)}, \wh\gamma_h^{(r)}, \wh\bzeta_g^{(r)},
\wh\bdelta_h^{(r)}, \wh\alpha_{gh}^{(r)}$ and the corresponding
estimated group number as $\wh G^{(r)}$ and $\wh H^{(r)}$.

\subsubsection{Estimation when $G = G_0, H=H_0$}

We first evaluate the estimation accuracy when  the group numbers are correctly specified.
Take $\blambda =(\lambda_1, \cdots, \lambda_G )^\top$ for example.
Denote $\wh \blambda^{(r)}$ as the estimator of $\blambda^0=
(\lambda_1^0, \cdots, \lambda_G^0 )^\top$ in the $r$th replicate.
To evaluate the estimation accuracy, we calculate the root mean squared
error (RMSE) as $\text{RMSE}_{\blambda} = \{R^{-1} \sum_{r = 1}^R
(\|\wh\blambda^{(r)} - \blambda_{0}\|^2) \}^{1/2}$.
Next, to gauge the performance of the statistical inference, we
construct the 95\% confidence interval for each parameter.
For example, denote the estimated standard error of $\lambda_g$ as
$\wh{\text{SE}}_{\lambda_g}^{(r)}$ for the $r$th replicate, then the
95\% confidence interval for $\wh\lambda_g^{(r)}$ is constructed as
$\text{CI}_{\lambda_g}^{(r)} = (\wh\lambda_g^{(r)} - 1.96 \times
\wh{\text{SE}}_{\lambda_g}^{(r)}, \wh\lambda_g^{(r)} +1.96 \times
\wh{\text{SE}}_{\lambda_g}^{(r)})$. Here
$\wh{\text{SE}}_{\lambda_g}^{(r)}$ is obtained by Theorem
\ref{thm:normal}. Subsequently, the coverage probability (CP) is
formed as $\text{CP}_{\lambda_g} = R^{-1} \sum_{r =1}^R
I(\lambda_{g,0} \in\text{CI}_{\lambda_g}^{(r)})$.
We calculate the CPs for other parameters similarly.
For comparison, we also calculate the RMSE and CP values for the oracle
estimators under the true group memberships (denoted as $\wh\lambda_{g}^{\text{or}},
\wh\gamma_{h}^{\text{or}}, \wh\bzeta_{g}^{\text{or}}, \wh\bdelta_{h}^{\text{or}}, \wh\alpha_{gh}^{\text{or}}$ accordingly).
Lastly, to evaluate the group memberships estimation, we calculate the
mis-clustering rates for the row and column groups as
$\wh\eta_1 = (N_1R)^{-1}\sum_{r =  1}^R \sum_i I(\wh g_i^{(r)}\ne
g_i^0)$ and $\wh\eta_2 = (N_2R)^{-1}\sum_{r =  1}^R \sum_j I(\wh
h_j^{(r)}\ne h_j^0)$, where $\wh g_i^{(r)}$ is the estimated group
membership of the $i$th node and $\wh h_j^{(r)}$ is defined similarly
for the $j$th node.
Here the mis-clustering rates are calculated after proper group permutations.

The simulation results are shown in Tables
\ref{table:sce1}--\ref{table:sce3} in Appendix \ref{subsec:appd_simu} under the three- different
combinations of group numbers.
The first finding across all combinations is that once the group
numbers are specified as the true values in advance, our iterative method
can estimate the true group memberships with high accuracy especially when the
sample size is large.
As $N_1,N_2$ or $T$ increase, the mis-clustering rates for either the row or the column groups reduce to around zero.
Furthermore, we note that the RMSEs decrease either when the network sizes
$N_1$ and $N_2$ increase or the time length $T$ increases, and they
approach the oracle RMSEs when the sample sizes  are
large.
Next, we inspect the statistical inference results. We observe that in Table
\ref{table:sce3}, the CPs are slightly small when the sample sizes are
not very large, but they grow up to around 0.95 as $N_1, N_2$ and $T$
increase.
This guarantees that even under the complex scenario where the number
of parameters to be estimated is large, our proposed method can
still perform well in terms of both estimation and inference for
sufficiently large sample sizes.
In addition, the CPs for all parameters are stable around 0.95 in Tables
\ref{table:sce1} and \ref{table:sce2} in the Appendix, which reflects the high
accuracy of the inference procedure.

\subsubsection{Estimation when $G \ge G_0, H\ge H_0$}

We next consider the case of estimation without specifying the true group numbers in advance.
Specifically, we estimate the group numbers by QIC in Section
\ref{sec:group_number}, where the tuning parameter is set to be $\kappa = 1/\{40 \log(T) T^{1/8}\}$.
Let the true group numbers be $G_0 = 3$ and $H_0 = 3$, and the
corresponding true parameters are shown in Table
\ref{table:true_param}.
To evaluate the estimation accuracy, we calculate the RMSE for each
parameter as explained below.
Take $\blambda$ for example, define $\text{RMSE}_{\blambda, all} =
\{(R N_1)^{-1}\sum_{r = 1}^R \sum_{i = 1}^{N_1} \|\wh\blambda_{\wh
	g_i}^{(r)} - \blambda_{g_i}\|^2\}^{1/2}$ as the RMSE for all
nodes. RMSE for other parameters are calculated similarly.
For the group memberships, the mis-clustering rates are calculated
following the idea of \cite{zhu2022simultaneous}.
Recall that we have $\wh \cR_g = \{i: \wh g_i  = g\}$ and $\wh \mC_h = \{j: \wh h_j = h\}$, where $\wh g_i$ and $\wh h_j$ are denoted as the estimated group membership for the $i$th and $j$th node.
Note that $G$ and $H$ are not necessarily equal to $G_0$ and $H_0$.
In this case, we define the mappings from the estimated group memberships to the true
group memberships $\chi_1: \{1,\cdots,G\} \to \{1, \cdots, G_0\}$ and
$\chi_2: \{1,\cdots,H\} \to \{1, \cdots, H_0\}$ as
\begin{align*}
	& \chi_1(g) = \argmax_{g' \in \{1, \cdots, G_0\}} \sum_{i = 1}^{N_1} I\big(i \in \wh\cR_g, g_i^0 = g'\big), ~~~ g \in \{1, \cdots, G\},\\
	& \chi_2(h) = \argmax_{h' \in \{1, \cdots, H_0\}} \sum_{i = 1}^{N_2} I\big(j \in \wh\mC_h, h_j^0 = h'\big), ~~~ h \in \{1, \cdots, H\}.
\end{align*}
Thus, the mapping $\chi_1(g)$ maps group $g$ to the true membership
$g'$ where the majority of nodes in $\wh \cR_g$ belong to.
Then, for the row group memberships, the mis-clustering rate in the
$r$th replicate is  defined as
\begin{align*}
	\wh\xi_1^{(r)} = N_1^{-1} \sum_{g=1}^G \sum_{i=1}^{N_1} I\big(i \in \wh\cR_g^{(r)}, g_i^0 \neq \chi_1(g) \big),
\end{align*}
where $\wh\cR_g^{(r)}$ is the
estimated node set belong to group
$g$ in the $r$th replicate.
We define the mis-clustering rate for the column group as
$\wh\xi_2^{(r)}$ similarly.
Then, the overall group memberships error rate is calculated as
$\wh\xi_1 = R^{-1} \sum_{r} \wh\xi_1^{(r)}$ and $\wh\xi_2 = R^{-1}
\sum_{r} \wh\xi_2^{(r)}$.
Additionally, to evaluate the performance of the group selection
criterion QIC, we define
\begin{align*}
	& \varrho(G) = R^{-1} \sum_{r=1}^R I(\wh G^{(r)} = G), \\
	& \varrho(H) = R^{-1} \sum_{r=1}^R I(\wh H^{(r)} = H).
\end{align*}
for each $G$ and $H$,
where $\wh G^{(r)}$ and $\wh H^{(r)}$ are the estimated group numbers
in the $r$th replicate.
Both $\varrho(G)$ and $\varrho(H)$ assess the proportion of the correctly
estimated group numbers for different $G$ and $H$,
and we wish to see a high ratio for $\varrho(G_0)$ and $\varrho(H_0)$.
The results are shown in Table \ref{table:QIC_1}--\ref{table:QIC_2} in Appendix \ref{subsec:appd_simu}.

We discuss the results in Table \ref{table:QIC_1} from two aspects.
On one hand, when the group number is under-specified ($G = 2, H =
2$), the node-wise RMSEs are large and usually do not decrease when the $N_1,
N_2$ and $T$ grow.
Besides, the error rates $\wh\xi_1$ and $\wh\xi_2$ are around 0.3,
indicating a low accuracy in estimating node memberships.
These results are expected since a non-ignorable estimation bias
exists in an under-fitted model.
On the other hand, when the group numbers $G$ and $H$ are correctly ($G=3,
H=3$) or over-specified ($G =4,  H = 4$), the RMSE values are
generally much lower.
This is consistent with our theoretical analysis in Theorem \ref{thm:pseudo_dist}.
From Table \ref{table:QIC_2}, we also observe that
the RMSEs decrease when $N_1, N_2$
and $T$ increase.
Moreover, both $\varrho(G)$ and $\varrho(H)$ tend to 1 for the
correct model ($G=3,  H= 3$) when the sample size is large.

\section{Real Data Applications}\label{sec:real}

We now proceed to apply our proposed method to two real-world cases. In the first case, we work with a dataset obtained from the Yelp official website (https://www.yelp.com/).  Our objective is to analyze user reviews pertaining to businesses and shops in different geographical districts.
In the second case, we collect data related to multilateral trade among countries from the International Monetary Fund (IMF) {\it Direction of Trade Statistics} (DOTS) \citep{imf2017}. The goal is to analyze international trade patterns over a specific time period. The second application can be found in Appendix \ref{subsec:trading}.


\subsection{Data Description}\label{sec:yelp}

The Yelp dataset spans from 2010 to 2018 and covers five North American cities: Charlotte, Las Vegas, Phoenix, Scottsdale, and Toronto. The observation period is divided into $T=36$ quarters. To ensure data quality, we filter the dataset to retain active users who have provided more than 5 reviews over this time span.
We further divide each city into districts, as illustrated in Figure
\ref{fig:map}. Table \ref{tbl:yelp_des} provides information on the
number of active users ($N_1$) and districts ($N_2$) in each city. Our
response variable, denoted as $Y_{ijt}$, represents the $\log(1+x)$ transformed number of
reviews by user $i$ on district $j$ during the $t$th quarter.
 Here $Y_{ijt}$ is treated as a continuous variable in our real data analysis.
To visualize the temporal trends in review activity, we calculate the quarterly average responses for each city, which are depicted in Figure \ref{fig:des_line}. Different patterns emerge from this analysis. For instance, Las Vegas stands out as the city with the highest  number of reviews, reflecting its bustling business environment. Charlotte, Phoenix, and Scottsdale exhibit relatively similar and stable review trends. In contrast, Toronto displays a noticeable increase in  review numbers after 2015, likely attributed to Yelp's expanded presence in the Toronto area during that time.

\begin{table}[]
	\centering
	\caption{Descriptive statistics of Yelp dataset.}
	\label{tbl:yelp_des}
	\scalebox{0.8}{
	\begin{tabular}{c|cc|cc}
		\hline
		City       & $N_1$ & $N_2$ & $d_{\text{user}}$ & $d_{\text{dis}}$ \\ \hline
		Charlotte  & 240  & 60    & 0.0030            & 0.0593            \\
		Las Vegas  & 826 & 64    & 0.0009            & 0.0590            \\
		Phoenix    & 323   & 63    & 0.0021            & 0.0589            \\
		Scottsdale & 391    & 60    & 0.0019            & 0.0599            \\
		Toronto    & 462    & 56    & 0.0016            & 0.0617           \\ \hline
	\end{tabular}
}
\end{table}

Next, we construct the adjacency matrices among the users ($\bA_1$) and districts ($\bA_2$) respectively as follows.
The user network is built based on the friend list information.
Specifically, if user $j$ is on the friend list of user $i$ on Yelp,
then we set $a_{1ij} = 1$. Otherwise we set $a_{1ij} = 0$.
The spatial network is built based on the geographical adjacent
relationship. Specifically,
we set $a_{2ij} = 1$ if the district $j$ is adjacent to district $i$.
We calculate the densities for the above two networks,
i.e., $d_{\text{user}} = \sum_i\sum_j a_{1ij}/\{N_1(N_1-1)\}$ and
$d_{\text{dis}} = \sum_i\sum_j a_{2ij}/\{N_2(N_2-1)\}$, which are
shown in Table \ref{tbl:yelp_des}. One could observe that the user
networks are quite sparse in all cities.

\begin{figure}[htpb!]
	\begin{center}
		\includegraphics[width=0.7\textwidth]{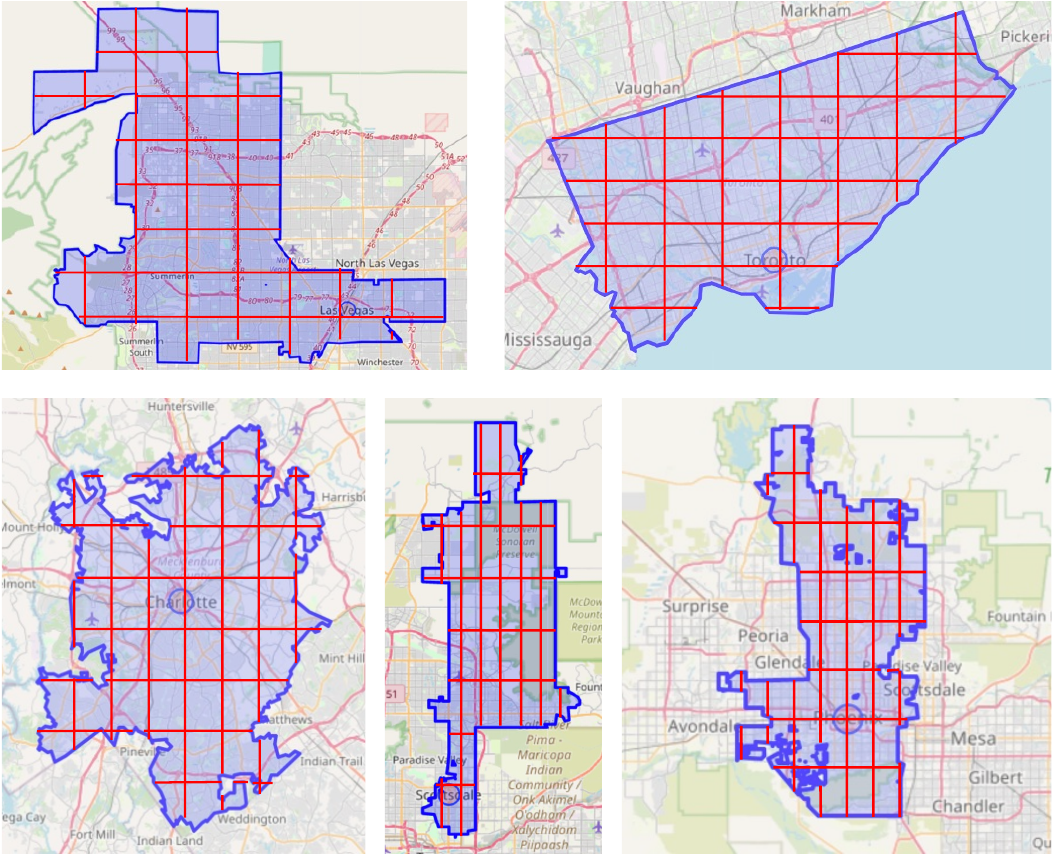}
		\caption{\small Geographical maps with split districts in
			each city. The city from top left panel to bottom
			right panel show the map of Las Vegas, Toronto,
			Charlotte, Scottsdale, and Phoenix, respectively.}
		\label{fig:map}
	\end{center}
\end{figure}

\begin{figure}[htpb!]
	\begin{center}
		\includegraphics[width=0.6\textwidth]{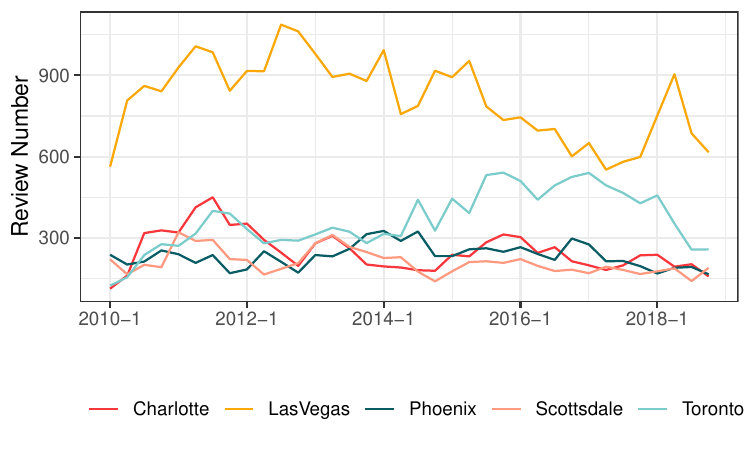}
		\caption{\small Average number of reviews from 2010-Q1 to 2018-Q4 in five cities.}
		\label{fig:des_line}
	\end{center}
\end{figure}

Lastly, to characterize the dynamic patterns of the responses, we
collect a number of covariates for users and districts, respectively.
For user $i$ in quarter $t$, we consider the following five
covariates: (1) the number of months after joining Yelp by
the start of the quarter $t$ ($x_{it, \text{dur}}$), (2) whether the
user is VIP by the start of the quarter $t$ ($x_{it, \text{vip}}$),
(3) average tags (i.e., ``useful'', ``funny'' and ``cool'') the user
$i$ obtains for his/her reviews during the last quarter ($x_{it,
	\text{use}}, x_{it, \text{fun}}, x_{it, \text{cool}}$).
Next,
for the $j$th district in quarter $t$, we consider two covariates: (1) the average ``stars'' ($z_{jt, \text{star}}$), and (2)
the average review number ($z_{jt, \text{num}}$) obtained by the $j$th
district during the $(t-1)$th quarter.
These two covariates are indicative of the average popularity levels in the preceding time period.
	We standardize all continuous covariates to be in the range $[0,1]$ for the subsequent analysis.
In Figure \ref{fig:box} in Appendix \ref{subsec:realdata}, we visualize the relationship between these user-related covariates and the response variable. The plot reveals that users who receive more tags for their reviews tend to be motivated to contribute more reviews themselves.
Notably, VIP users in Scottsdale and Toronto tend to write more reviews, whereas VIP users in Charlotte exhibit comparatively less activity. Subsequently, we apply the proposed GMNAR model independently to each of the five cities, enabling us to analyze and understand the distinctive group patterns within each urban area.

\subsection{Estimation Results}\label{subsec:yelp_res}

We employ  QIC for the selection of group numbers, and the estimation results are detailed in Table \ref{tbl:real_res_1} and Table \ref{tbl:real_res_2} in Appendix \ref{subsec:realdata}. The numbers of user groups and district groups vary across the five cities, indicating different levels of heterogeneity among them. For instance, consider the results for Phoenix. Notably, the estimated $\wh\balpha$ values are all positive, indicating a positive self-motivated effect. Additionally, the spatial (column) network effects are consistently positive, suggesting a favorable effect from neighboring districts. Such an observation is consistent with the findings in the literature \cite[e.g.][]{sun2017spatial}. Furthermore, we can also observe that within these spatial effects, $\wh\gamma_2 = 0.270$ is the largest, signifying a strong neighbor effect.
{In other words, if the second group districts' neighbors obtain more reviews in the last period, then it is likely that these district would receive more reviews in this period.}
This implies that the districts in this group are closely
interconnected to their neighboring counterparts.
We further visualize the districts by estimated groups in the left
panel of Figure \ref{fig:dis_block_map}, where the pink
districts are from the first group and the green ones are from the
second one. We can observe obvious  blocks (marked as red) in
  group 2. To
further investigate the local area of the three main blocks
shown in the figure, we map some of the shops in these districts to the Google Map, and mark some restaurants on it (as shown in the right panel). The first block is concentrated around the central city area, and most shops are located in downtown, where large commercial and government buildings are placed.
The second block is in/near the Camelback East Village. This area has
two main  streets, and most of the shops are located in
these two streets.
The third block is in the northern region, spreading from the North Mountain Village to the Deer Valley area, which include some mountains and hills.
The three blocks, as shown in the Wikipedia of Phoenix, have their own features, which may lead to three business patterns, within each of which the connections are quite close due to the inner similarity.
\begin{figure}[htpb!]
	\begin{center}
		\includegraphics[width=0.7\textwidth]{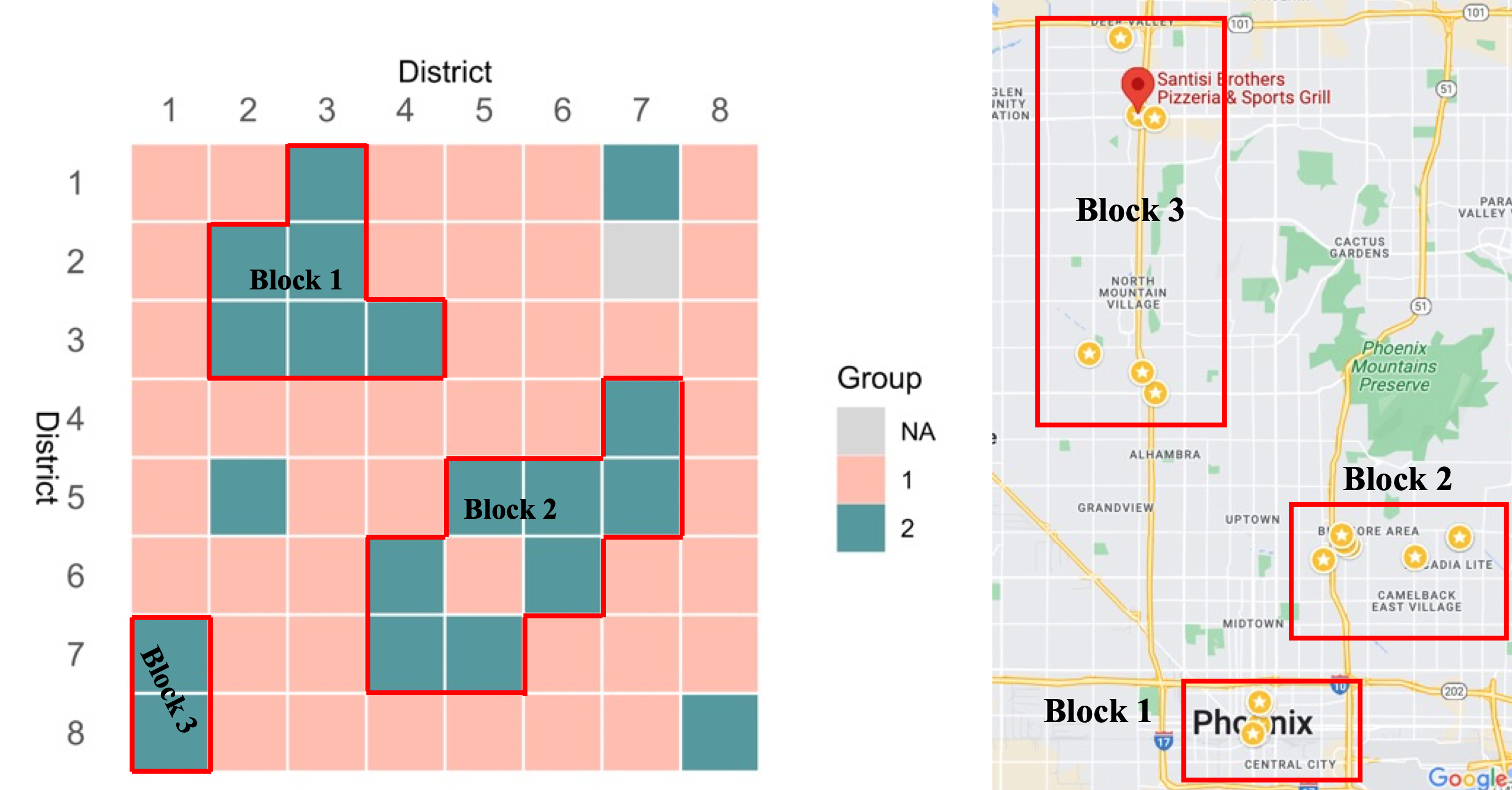}
		\caption{\small The left panel shows the 63 districts in Phoenix and their group memberships are visualized by different color. The gray district has no shops in the data. In the second group, three main blocks are highlighted by red lines. The right panel shows several shops in the three blocks on Google Map, marked by yellow stars. Note that the larger longitude (larger district number) means norther area.}
		\label{fig:dis_block_map}
	\end{center}
\end{figure}
Performing similar investigation on the user dimension, one
can find that the social (row) network effect in the first group
appears to be significantly negative ($\wh\lambda_1 = -0.02$),
whereas corresponding coefficients in the other two groups are
positive. This implies that user activities in the first group are
influenced oppositely by their friends' behaviors, whereas in the
other groups, users are still positively influenced by their
friends. Considering that the social network effect is generally found
to be positive on the choice of restaurant
\citep{tiwari2016social,fe2023social}, it is of interest to look
closer into this user subgroup and explain why they react negatively
toward actions of friends they follow.

Further,
we discuss how the response relates to user and district level covariates. Firstly, concerning user-related covariates, we observe that users who receive more ``useful'' tags for their reviews in the last quarter tend to write more reviews in the subsequent quarter. This reflects an encouraging effect on user behavior, indicating that users may become more active when their comments are valued as ``valuable'' by others. Secondly, regarding district-related covariates, we find that districts with higher star ratings in the last quarter tend to attract more customers on this platform, subsequently leading to an increase in the number of reviews in the next quarter. This underscores the positive impact of high ratings on customer engagement and review generation.

\begin{sidewaystable}[]
	\centering
	\caption{\small Estimation results for Charlotte, Las Vegas and Phoenix. The $p$-values are shown in the parenthesis.
Take Phoenix for example, $\lambda$ is clustered into three groups, with $\lambda_g$ representing the social network effect on the users in the $g$th group from their friends' behavior.
	$\bgamma$ is clustered into two groups, with $\gamma_h$ meaning the spatial network effect on the districts in the $h$th group from their adjacent neighbors.
$\bxi$ and $\bdelta$ are the covariates effects, and we take $\xi_{\textbf{use}}$ to illustrate. It is also been clustered into three groups, with the coefficient in the $g$th group meaning the effect on the users in the $g$th group from the average ``useful'' tag.
$\alpha$ is clustered into three groups by row and into two groups by column, where $\alpha_{gh}$ means the self-momentum effect of the users in the $g$th group by row and $h$th group by column.
}
	\label{tbl:real_res_1}
	\scalebox{0.65}{
			\begin{tabular}{c|cccccc|ccccc|ccccc}
				\hline
				Parameters                                    & \multicolumn{6}{c|}{Charlotte}                                                                                                                                                                                                                                                                                                                                                                                                                      & \multicolumn{5}{c|}{LasVegas}                                                                                                                                                                                                                                                                                                                                                  & \multicolumn{5}{c}{Phoenix}                                                                                                                                                                                                                                                                                                                                                    \\ \hline
				\multirow{2}{*}{}                             & \multicolumn{3}{c|}{$\lambda_g$}                                                                                                                                                                                                    & \multicolumn{3}{c|}{$\gamma_h$}                                                                                                                                                                               & \multicolumn{3}{c|}{$\lambda_g$}                                                                                                                                                                                                     & \multicolumn{2}{c|}{$\gamma_h$}                                                                                                         & \multicolumn{3}{c|}{$\lambda_g$}                                                                                                                                                                                                     & \multicolumn{2}{c}{$\gamma_h$}                                                                                                          \\
				& \begin{tabular}[c]{@{}c@{}}0.004\\ (0.017)\end{tabular}            & \begin{tabular}[c]{@{}c@{}}0.110\\ (\textless{}0.001)\end{tabular}  & \multicolumn{1}{c|}{\begin{tabular}[c]{@{}c@{}}0.047\\ (\textless{}0.001)\end{tabular}}  & \begin{tabular}[c]{@{}c@{}}0.035\\ (\textless{}0.001)\end{tabular}  & \begin{tabular}[c]{@{}c@{}}0.475\\ (\textless{}0.001)\end{tabular} & \begin{tabular}[c]{@{}c@{}}0.286\\ (\textless{}0.001)\end{tabular} & \begin{tabular}[c]{@{}c@{}}0.068\\ (\textless{}0.001)\end{tabular}  & \begin{tabular}[c]{@{}c@{}}0.008\\ (\textless{}0.001)\end{tabular}  & \multicolumn{1}{c|}{\begin{tabular}[c]{@{}c@{}}0.054\\ (\textless{}0.001)\end{tabular}}  & \begin{tabular}[c]{@{}c@{}}0.150\\ (\textless{}0.001)\end{tabular} & \begin{tabular}[c]{@{}c@{}}0.326\\ (\textless{}0.001)\end{tabular} & \begin{tabular}[c]{@{}c@{}}-0.020\\ (0.101)\end{tabular}            & \begin{tabular}[c]{@{}c@{}}0.007\\ (\textless{}0.001)\end{tabular}  & \multicolumn{1}{c|}{\begin{tabular}[c]{@{}c@{}}0.024\\ (\textless{}0.001)\end{tabular}}  & \begin{tabular}[c]{@{}c@{}}0.050\\ (\textless{}0.001)\end{tabular} & \begin{tabular}[c]{@{}c@{}}0.270\\ (\textless{}0.001)\end{tabular} \\ \hline
				& \multicolumn{3}{c|}{$\bzeta_g$}                                                                                                                                                                                                     & \multicolumn{3}{c|}{$\bdelta_h$}                                                                                                                                                                              & \multicolumn{3}{c|}{$\bzeta_g$}                                                                                                                                                                                                      & \multicolumn{2}{c|}{$\bdelta_h$}                                                                                                        & \multicolumn{3}{c|}{$\bzeta_g$}                                                                                                                                                                                                      & \multicolumn{2}{c}{$\bdelta_h$}                                                                                                         \\
				Intercept                                     & \begin{tabular}[c]{@{}c@{}}0.001\\ (0.328)\end{tabular}            & \begin{tabular}[c]{@{}c@{}}-0.004\\ (\textless{}0.001)\end{tabular}  & \multicolumn{1}{c|}{\begin{tabular}[c]{@{}c@{}}0.003\\ (0.007)\end{tabular}}             & \begin{tabular}[c]{@{}c@{}}-0.004\\ (\textless{}0.001)\end{tabular} & \begin{tabular}[c]{@{}c@{}}0.041\\ (\textless{}0.001)\end{tabular} & \begin{tabular}[c]{@{}c@{}}$-10^{-4}$ \\ (0.796)\end{tabular}           & \begin{tabular}[c]{@{}c@{}}0.011\\ (\textless{}0.001)\end{tabular}  & \begin{tabular}[c]{@{}c@{}}-0.005\\ (\textless{}0.001)\end{tabular} & \multicolumn{1}{c|}{\begin{tabular}[c]{@{}c@{}}-0.006\\ (0.586)\end{tabular}}             & \begin{tabular}[c]{@{}c@{}}0.010\\ (\textless{}0.001)\end{tabular} & \begin{tabular}[c]{@{}c@{}}0.003\\ (0.011)\end{tabular}            & \begin{tabular}[c]{@{}c@{}}0.058\\ (\textless{}0.001)\end{tabular}  & \begin{tabular}[c]{@{}c@{}}-0.052\\ (\textless{}0.001)\end{tabular} & \multicolumn{1}{c|}{\begin{tabular}[c]{@{}c@{}}-0.006\\ (\textless{}0.001)\end{tabular}} & \begin{tabular}[c]{@{}c@{}}0.050\\ (\textless{}0.001)\end{tabular} & \begin{tabular}[c]{@{}c@{}}0.021\\ (\textless{}0.001)\end{tabular} \\
				$\zeta_{\text{dur}}$ / $\delta_{\text{star}}$ & \begin{tabular}[c]{@{}c@{}}-0.001\\ (0.162)\end{tabular}           & \begin{tabular}[c]{@{}c@{}}0.050\\ (\textless{}0.001)\end{tabular}  & \multicolumn{1}{c|}{\begin{tabular}[c]{@{}c@{}}0.015\\ (\textless{}0.001)\end{tabular}}  & \begin{tabular}[c]{@{}c@{}}0.002\\ (\textless{}0.001)\end{tabular}  & \begin{tabular}[c]{@{}c@{}}-0.012\\ (0.078)\end{tabular}           & \begin{tabular}[c]{@{}c@{}}0.004\\ (0.005)\end{tabular}            & \begin{tabular}[c]{@{}c@{}}0.059\\ (\textless{}0.001)\end{tabular}  & \begin{tabular}[c]{@{}c@{}}0.001\\ (0.194)\end{tabular}             & \multicolumn{1}{c|}{\begin{tabular}[c]{@{}c@{}}0.019\\ (\textless{}0.001)\end{tabular}}  & \begin{tabular}[c]{@{}c@{}}$10^{-4}$\\ (0.170)\end{tabular}            & \begin{tabular}[c]{@{}c@{}}0.001\\ (0.744)\end{tabular}            & \begin{tabular}[c]{@{}c@{}}-0.093\\ (\textless{}0.001)\end{tabular} & \begin{tabular}[c]{@{}c@{}}0.001\\ (0.150)\end{tabular}             & \multicolumn{1}{c|}{\begin{tabular}[c]{@{}c@{}}0.002\\ (0.148)\end{tabular}}             & \begin{tabular}[c]{@{}c@{}}0.003\\ (\textless{}0.001)\end{tabular} & \begin{tabular}[c]{@{}c@{}}0.004\\ (0.010)\end{tabular}            \\
				$\zeta_{\text{vip}}$ / $\delta_{\text{num}}$  & \begin{tabular}[c]{@{}c@{}}0.001\\ (0.012)\end{tabular}            & \begin{tabular}[c]{@{}c@{}}-0.037\\ (\textless{}0.001)\end{tabular} & \multicolumn{1}{c|}{\begin{tabular}[c]{@{}c@{}}-0.002\\ (0.044)\end{tabular}}            & \begin{tabular}[c]{@{}c@{}}0.007\\ (\textless{}0.001)\end{tabular}  & \begin{tabular}[c]{@{}c@{}}0.008\\ (0.001)\end{tabular}            & \begin{tabular}[c]{@{}c@{}}0.003\\ (0.233)\end{tabular}            & \begin{tabular}[c]{@{}c@{}}0.028\\ (\textless{}0.001)\end{tabular}  & \begin{tabular}[c]{@{}c@{}}0.000\\ (0.093)\end{tabular}             & \multicolumn{1}{c|}{\begin{tabular}[c]{@{}c@{}}-0.010\\ (\textless{}0.001)\end{tabular}} & \begin{tabular}[c]{@{}c@{}}0.016\\ (\textless{}0.001)\end{tabular} & \begin{tabular}[c]{@{}c@{}}0.038\\ (\textless{}0.001)\end{tabular} & \begin{tabular}[c]{@{}c@{}}0.008\\ (\textless{}0.001)\end{tabular}  & \begin{tabular}[c]{@{}c@{}}$10^{-4}$\\ (0.054)\end{tabular}             & \multicolumn{1}{c|}{\begin{tabular}[c]{@{}c@{}}$10^{-4}$\\ (0.369)\end{tabular}}             & \begin{tabular}[c]{@{}c@{}}0.002\\ (\textless{}0.001)\end{tabular} & \begin{tabular}[c]{@{}c@{}}$10^{-4}$\\ (0.938)\end{tabular}            \\
				$\zeta_{\text{use}}$                          & \begin{tabular}[c]{@{}c@{}}0.017\\ (\textless{}0.001)\end{tabular} & \begin{tabular}[c]{@{}c@{}}0.104\\ (\textless{}0.001)\end{tabular}  & \multicolumn{1}{c|}{\begin{tabular}[c]{@{}c@{}}0.052\\ (\textless{}0.001)\end{tabular}}  &                                                                     &                                                                    &                                                                    & \begin{tabular}[c]{@{}c@{}}1.115\\ (\textless{}0.001)\end{tabular}  & \begin{tabular}[c]{@{}c@{}}0.011\\ (\textless{}0.001)\end{tabular}  & \multicolumn{1}{c|}{\begin{tabular}[c]{@{}c@{}}0.143\\ (\textless{}0.001)\end{tabular}}  &                                                                    &                                                                    & \begin{tabular}[c]{@{}c@{}}0.385\\ (\textless{}0.001)\end{tabular}  & \begin{tabular}[c]{@{}c@{}}0.011\\ (\textless{}0.001)\end{tabular}  & \multicolumn{1}{c|}{\begin{tabular}[c]{@{}c@{}}0.071\\ (\textless{}0.001)\end{tabular}}  &                                                                    &                                                                    \\
				$\zeta_{\text{fun}}$                          & \begin{tabular}[c]{@{}c@{}}-0.001\\ (0.806)\end{tabular}           & \begin{tabular}[c]{@{}c@{}}-0.020\\ (0.002)\end{tabular}            & \multicolumn{1}{c|}{\begin{tabular}[c]{@{}c@{}}$10^{-4}$\\ (0.995)\end{tabular}}             &                                                                     &                                                                    &                                                                    & \begin{tabular}[c]{@{}c@{}}-0.589\\ (\textless{}0.001)\end{tabular} & \begin{tabular}[c]{@{}c@{}}-0.001\\ (0.795)\end{tabular}            & \multicolumn{1}{c|}{\begin{tabular}[c]{@{}c@{}}-0.289\\ (\textless{}0.001)\end{tabular}} &                                                                    &                                                                    & \begin{tabular}[c]{@{}c@{}}0.064\\ (0.080)\end{tabular}             & \begin{tabular}[c]{@{}c@{}}-0.008\\ (0.002)\end{tabular}            & \multicolumn{1}{c|}{\begin{tabular}[c]{@{}c@{}}-0.033\\ (\textless{}0.001)\end{tabular}} &                                                                    &                                                                    \\
				$\zeta_{\text{cool}}$                         & \begin{tabular}[c]{@{}c@{}}-0.001\\ (0.908)\end{tabular}           & \begin{tabular}[c]{@{}c@{}}0.003\\ (0.621)\end{tabular}             & \multicolumn{1}{c|}{\begin{tabular}[c]{@{}c@{}}-0.017\\ (\textless{}0.001)\end{tabular}} &                                                                     &                                                                    &                                                                    & \begin{tabular}[c]{@{}c@{}}-0.644\\ (\textless{}0.001)\end{tabular} & \begin{tabular}[c]{@{}c@{}}-0.008\\ (0.043)\end{tabular}            & \multicolumn{1}{c|}{\begin{tabular}[c]{@{}c@{}}0.217\\ (\textless{}0.001)\end{tabular}}  &                                                                    &                                                                    & \begin{tabular}[c]{@{}c@{}}0.208\\ (\textless{}0.001)\end{tabular}  & \begin{tabular}[c]{@{}c@{}}0.010\\ (0.005)\end{tabular}             & \multicolumn{1}{c|}{\begin{tabular}[c]{@{}c@{}}-0.011\\ (0.007)\end{tabular}}            &                                                                    &                                                                    \\ \hline
				\multirow{4}{*}{}                             & \multicolumn{6}{c|}{$\balpha \in \mR^{G \times H}$}                                                                                                                                                                                                                                                                                                                                                                                                 & \multicolumn{5}{c|}{$\balpha \in \mR^{G \times H}$}                                                                                                                                                                                                                                                                                                                            & \multicolumn{5}{c}{$\balpha \in \mR^{G \times H}$}                                                                                                                                                                                                                                                                                                                             \\
				& \multicolumn{2}{c}{\begin{tabular}[c]{@{}c@{}}0.019\\ (0.002)\end{tabular}}                                                              & \multicolumn{2}{c}{\begin{tabular}[c]{@{}c@{}}0.051\\ (\textless{}0.001)\end{tabular}}                                                                         & \multicolumn{2}{c|}{\begin{tabular}[c]{@{}c@{}}0.058\\ (\textless{}0.001)\end{tabular}}                                                 & \multicolumn{2}{c}{\begin{tabular}[c]{@{}c@{}}0.288\\ (\textless{}0.001)\end{tabular}}                                                    & \begin{tabular}[c]{@{}c@{}}0.021\\ (\textless{}0.001)\end{tabular}                       & \multicolumn{2}{c|}{\begin{tabular}[c]{@{}c@{}}0.174\\ (\textless{}0.001)\end{tabular}}                                                 & \multicolumn{2}{c}{\begin{tabular}[c]{@{}c@{}}0.026\\ (\textless{}0.001)\end{tabular}}                                                    & \begin{tabular}[c]{@{}c@{}}0.026\\ (\textless{}0.001)\end{tabular}                       & \multicolumn{2}{c}{\begin{tabular}[c]{@{}c@{}}0.047\\ (\textless{}0.001)\end{tabular}}                                                  \\
				& \multicolumn{2}{c}{\begin{tabular}[c]{@{}c@{}}0.005\\ (0.218)\end{tabular}}                                                              & \multicolumn{2}{c}{\begin{tabular}[c]{@{}c@{}}0.374\\ (\textless{}0.001)\end{tabular}}                                                                         & \multicolumn{2}{c|}{\begin{tabular}[c]{@{}c@{}}0.224\\ (\textless{}0.001)\end{tabular}}                                                 & \multicolumn{2}{c}{\begin{tabular}[c]{@{}c@{}}0.478\\ (\textless{}0.001)\end{tabular}}                                                    & \begin{tabular}[c]{@{}c@{}}0.036\\ (\textless{}0.001)\end{tabular}                       & \multicolumn{2}{c|}{\begin{tabular}[c]{@{}c@{}}0.330\\ (\textless{}0.001)\end{tabular}}                                                 & \multicolumn{2}{c}{\begin{tabular}[c]{@{}c@{}}0.193\\ (\textless{}0.001)\end{tabular}}                                                    & \begin{tabular}[c]{@{}c@{}}0.021\\ (\textless{}0.001)\end{tabular}                       & \multicolumn{2}{c}{\begin{tabular}[c]{@{}c@{}}0.146\\ (\textless{}0.001)\end{tabular}}                                                  \\
				& \multicolumn{2}{c}{\begin{tabular}[c]{@{}c@{}}0.017\\ (\textless{}0.001)\end{tabular}}                                                   & \multicolumn{2}{c}{\begin{tabular}[c]{@{}c@{}}0.200\\ (\textless{}0.001)\end{tabular}}                                                                         & \multicolumn{2}{c|}{\begin{tabular}[c]{@{}c@{}}0.115\\ (\textless{}0.001)\end{tabular}}                                                 & \multicolumn{2}{c}{}                                                                                                                      &                                                                                          & \multicolumn{2}{c|}{}                                                                                                                   & \multicolumn{2}{c}{}                                                                                                                      &                                                                                          & \multicolumn{2}{c}{}                                                                                                                    \\ \hline
			\end{tabular}
		}
	\end{sidewaystable}

	\section{Concluding Remarks}\label{sec:conclusion}

In this work, we introduce a novel Group Matrix Network Autoregression
(GMNAR) model specifically designed for time series data indexed by
two distinct heterogeneous networks. By leveraging network structures
on both rows and columns, our model establishes a unique framework for
analyzing matrix-valued time series data. Our proposed model is highly
interpretable and accommodates various types of network effects while
accounting for network heterogeneity. While we have focused on the case of two networks in our presentation, extending this framework to accommodate multiple networks is straightforward. Suppose there are $q$ networks characterized by adjacency matrices $\A_1,\cdots,\A_q$, each with its own latent group structure $\mG_l=(g_{i}^{(l)}:1\le i\le N_l)$ for its $N_l$ network nodes, where $1\le l\le q$. In such a scenario, we deal with a tensor-valued time series denoted as $\Y_t=(Y_{i_1i_2...i_q,t})\in \mR^{N_1\times N_2\times\cdots\times N_q}$. The model for this tensor-valued time series can be expressed as follows:
\[
\begin{split}
	Y_{i_1i_2...i_q,t} =\sum_{l=1}^q\underbrace{\lambda_{g_{i_l}^{(l)}}^{(l)}\sum_{k = 1}^{N_l} \frac{a_{li_{l}k}}{n_{li_l}}Y_{i_1...i_{l-1}ki_{l+1}...i_q,(t-1)}}_{\text{\rm The $l$th Network main effect}}
	&+ \underbrace{\alpha_{g_{i_l}^{(l)}...g_{i_q}^{(q)}}Y_{i_1i_2...i_q,(t-1)} }_{\text{\rm Self-momentum}}
	\\&+\sum_{l=1}^q\underbrace{\bx_{i_lt}^{(l)\top}\bzeta_{g_{i_l}}^{(l)}}_{\text{\rm The $l$th covariate effects}}+\ve_{i_1i_2...i_q,t},
\end{split}
\]
Here, $n_{li_l}=\sum_{k=1}^{N_1}a_{li_lk}$, $\bx_{i_lt}^{(l)}\in \mR^{p_l}$ represents exogenous covariates associated with the $i_l$th member in the $l$th network, and $\ve_{i_1i_2...i_q,t}$ denotes the white noise. Estimating this tensor-valued time series is feasible using a similar algorithm to our proposed Algorithm~\ref{alg:gmnar}, and the theoretical properties of the estimators can be established utilizing the technical tools developed in this work.
This model presents a valuable tool for analyzing data collected in complex network environments, shedding light on various network effects.
Furthermore, it is worth exploring models that consider row and column network effects in a multiplicative form, as investigated by \cite{chen2021autoregressive}.
Besides, introducing a hidden factor structure into the GMNAR model can potentially offer more insights into the high-dimensional data, capturing more underlying information within the matrix-valued time series, and thus will be an interesting future research topic.
Lastly, it is also important to further investigate how to model categorical responses in our modelling framework.

\bibliographystyle{asa}
\bibliography{./xuening}

\end{document}